\documentclass[12pt]{article}

\newfont{\bbold}{msbm10 scaled\magstep1}

\newcommand{\ro}{\varrho}

\newcommand{\beqa}{\begin{eqnarray}}
\newcommand{\eeqa}{\end{eqnarray}}
\newcommand{\beq}{\begin{equation}}
\newcommand{\eeq}{\end{equation}}
\newcommand{\rec}[1]{\mbox{$\frac{1}{#1}$}}
\newcommand{\mfrac}[2]{\mbox{$\frac{#1}{#2}$}}
\newcommand{\half}{\mbox{$\frac{1}{2}$}}

\newcommand{\nn}{\nonumber}

\newcommand{\etal}{{\em et al.}}

\newcommand{\dd}{{\rm d}}

\newcommand{\de}{\delta}
\newcommand{\ep}{\varepsilon}
\newcommand{\ds}{\displaystyle}

\let\tilde=\widetilde

\newcommand{\lbl}[1]{\label{#1}}

\renewcommand{\theequation}{\arabic{section}.\arabic{equation}}

\begin{document}

\title{{\bf Mathematical models of homochiralisation by grinding of crystals}}
\author{Jonathan AD Wattis \\ 
Theoretical Mechanics, School of Mathematical Sciences, \\
University of Nottingham, University Park, Nottingham NG7 2RD, UK. \\ 
Jonathan.Wattis@nottingham.ac.uk } 
\date{{\footnotesize\today}}

\maketitle

%-------------------------------------------------------

\begin{abstract}
We review the existing mathematical models which describe 
physicochemical mechanisms capable of producing a 
symmetry-breaking transition to a state in which one chirality 
dominates the other.  A new model is proposed, with the aim of 
elucidating the fundamental processes at work in the crystal 
grinding systems of Viedma [Phys Rev Lett 94, 065504, (2005)] and 
Noorduin [J Am Chem Soc 130, 1158, (2008)].   We simplify the 
model as far as possible to uncover the fundamental competitive 
process which causes the symmetry-breaking, and analyse other 
simplifications which might be  expected to show symmetry-breaking. 
\end{abstract}

%---------------------------------------------------------
\section{Introduction}
\label{intro-sec}
\setcounter{equation}{0}

A significant stage in the formation of living systems was the 
transition from a symmetric chemistry involving mirror-symmetric and 
approximately equal numbers of left- and right-handed chiral species 
into a system involving just one-handedness of chiral molecules.  

In this paper we focus on mathematical models of one example of a 
physicochemical system which undergoes such a symmetry-breaking 
transition, namely the crystal grinding processes investigated by 
Viedma \cite{viedma} and Noorduin \etal\  \cite{wim}, 
which have been recently reviewed by McBride \& Tully 
\cite{mcbride-nature}. Our aim is to describe this process by way 
of a detailed microscopic model of the nucleation and growth processes 
and then to simplify the model, retaining only the bare essential 
mechanisms responsible for the symmetry-breaking bifurcation. 

We start by reviewing the processes which are already known to 
cause a symmetry-breaking bifurcation. By this we mean that a system 
which starts off in a racemic state (one in which both left-handed and 
right-handed structures occur with approximately equal frequencies) 
and, as the system evolves, the two handednesses grow differently, 
so that at a later time, one handedness is predominant in the system.  

%---------------------------------------------------------
\subsection{Models for homochiralisation}

Many models have been proposed for the emergence of homochirality 
from an initially racemic mixture of precursors.   Frank \cite{frank} 
proposed an open system into which $R$ and $S$ particles are 
continually introduced, and combine to form one of two possible 
products: left-  or right-handed species, $X,Y$.   Each of these 
products acts as a catalyst for its own production (autocatalysis), and 
each combines with the opposing handed product (cross-inhibition) to 
form an inert product ($P$) which is removed from the system at some 
rate.   These processes are summarised by the following reaction scheme: 
\beq \begin{array}{rclcrclcl} 
&&&& \hspace*{-9mm}{\rm external \;\;\; source} & \rightarrow &R,S& 
	\;\; & {\rm input}, k_0, \\ 
R+S & \rightleftharpoons & X && 
R+S & \rightleftharpoons & Y &\qquad &\mbox{slow}, k_1 , \\ 
R+S+X & \rightleftharpoons & 2 X && 
R+S+Y & \rightleftharpoons & 2 Y &\quad& \mbox{fast, autocatalytic}, k_2 \\ 
&&&&X + Y & \rightarrow & P  &\qquad& \mbox{cross-inhibition}, k_3 , \\
&&&& P &\rightarrow & & \qquad & {\rm removal}, k_4 . 
\end{array}\eeq 
Ignoring the reversible reactions (for simplicity), 
this system can be modelled by the differential equations 
\beqa
\frac{\dd r}{\dd t} & =& k_0 - 2 k_1 r s - k_2 r s (x\!+\!y) 
	+ k_{-1} (x\!+\!y) + k_{-2} (x^2\!+\!y^2) , \\
\frac{\dd s}{\dd t} & = & k_0 - 2 k_1 r s - k_2 r s (x\!+\!y) 
	+ k_{-1} (x\!+\!y) + k_{-2} (x^2\!+\!y^2) , \\ 
\frac{\dd x}{\dd t} & = & 
	k_1 r s + k_2 r s x - k_3 x y - k_{-1} x - k_{-2} x^2 , \\ 
\frac{\dd y}{\dd t} & = & 
	k_1 r s + k_2 r s y - k_3 x y - k_{-1} y - k_{-2} y^2 , \\ 
\frac{\dd p}{\dd t} & = & k_{3} x y - k_4 p , 
\eeqa
from which we note that at steady-state we have 
\beq
rs=\frac{k_0+k_{-1}(x+y) + k_{-1}(x^2+y^2)}{2k_1+k_2(x+y)}.
\eeq
We write the absolute enantiomeric excess as $ee=x-y$ and the 
total concentration as $\sigma=x+y$;  adding and subtracting 
the equations for $\dd x/\dd t$ and $\dd y/\dd t$, we find 
\beq 
\sigma^2 = \frac{2k_0}{k_3} + ee^2 , 
\eeq 
\beq 
ee \left[ 
\frac{k_2(k_{-2}ee^2+k_{-2}\sigma^2+2k_{-1}\sigma+2k_0)}
{2(2k_1+k_2\sigma)} - k_{-1} - k_{-2} \sigma \right] = 0 . 
\eeq 
Hence $ee=0$ is always a solution, and there are other solutions with 
$ee\neq0$ if the rate constants $k_*$ satisfy certain conditions (these 
include $k_3>k_{-2}$ and $k_0$ being sufficiently large). 

The important issues to note here are: 
\begin{description}
\item[(i)] this system is {\em open}, it requires the continual supply of 
fresh $R,S$ to maintain the asymmetric steady-state. Also, the 
removal of products is required to avoid the input terms 
causing the total amount of material to increase indefinitely;  
\item[(ii)] the forcing input term drives the system away from 
an equilibrium solution, into a distinct steady-state solution;  
\item[(iii)] the system has cross-inhibition which removes equal 
numbers of $X$ and $Y$, amplifying any differences caused by 
random fluctuations in the initial data or in the input rates. 
\end{description} 

Saito \& Hyuga \cite{saito} discuss a sequence of toy models 
describing homochirality caused by nonlinear autocatalysis and 
recycling.  Their family of models can be summarised by 
\beqa
\frac{\dd r}{\dd t} & =& k r^2 (1-r-s) - \lambda r , \\ 
\frac{\dd s}{\dd t} & =& k s^2 (1-r-s) - \lambda s , 
\eeqa
where $r$ and $s$ are the concentrations of the two enantiomers. 
Initially they consider $k_r=k_s=k$ and $\lambda=0$ and find that 
enantiomeric exess, $r-s$ is constant.   Next the case $k_r=kr$, 
$k_s=ks$, $\lambda=0$ is analysed, wherein the relative enantiomeric 
excess $\frac{r-s}{r+s}$ is constant.  Then the more complex case of 
$k_r=k r^2$, $k_s=k s^2$, $\lambda=0$ is analysed, and amplification 
of the enantiomeric excess is obtained.  This amplification persists 
when the case $\lambda>0$ is finally analysed.  This shows us strong 
autocatalysis may cause homochiralisation, but in any given 
experiment, it is not clear which form of rate coefficients 
($k_r,k_s,\lambda$) should be used. 

Saito \& Hyuga (2005) analyse a series of models of crystallisation 
which include some of features present in our more general model.  
They note that a model truncated at tetramers exhibits different 
behaviour from one truncated at hexamers.  In particular, the 
symmetry-breaking phenomena is {\em not} present in the tetramer 
model, but {\em is} exhibited by the hexamer model.  Hence, later, 
we will consider models truncated at the tetramer and the hexamer 
levels and investigate the differences in symmetry-breaking 
behaviour (Sections \ref{tetra-sec} and \ref{hex-sec}).  

Denoting monomers by $c$, small and large left-handed clusters by 
$x_1,x_2$ respectively and right-handed by $y_1,y_2$, Uwaha 
\cite{uwaha} writes down the scheme 
\beqa
\frac{\dd c}{\dd t} & =& - 2 k_0 z^2 k_1 z (x_1+y_1) + 
	\lambda_1(x_2+y_2) + \lambda_0(x_1+y_1) , 
\\ 
\frac{\dd x_1}{\dd t} & = & k_0 z^2 - k_u x_1 x_2 
	- k_c x_1^2 + \lambda_u x_2 + \lambda_0 x_1 , 
\\ 
\frac{\dd x_2}{\dd t} & =& k_1 x_2 c + k_u x_1 x_2 + 
	k_c x_1^2 - \lambda_1 x_2 - \lambda_u x_2 , 
\\ 
\frac{\dd y_1}{\dd t} & = & k_0 z^2 - k_u y_1 y_2 
	- k_c y_1^2 + \lambda_u y_2 + \lambda_0 y_1 , 
\\ 
\frac{\dd y_2}{\dd t} & =& k_1 y_2 c + k_u y_1 y_2 + 
	k_c y_1^2 - \lambda_1 y_2 - \lambda_u y_2 , 
\eeqa
which models 
\begin{itemize}
\item the formation of small chiral clusters 
	($x_1,y_1$) from an achiral monomer ($c$) at rate $k_0$, 
\item small chiral clusters ($x_1,y_1$) of the same handedness 
	combining to form larger chiral clusters (rate $k_c$), 
\item small and larger clusters combining to form larger clusters 
(rate $k_u$), 
\item large clusters combining with achiral monomers to form more 
	large clusters at the rate $k_1$, 
\item the break up of larger clusters into smaller clusters (rate $\lambda_u$), 
\item the break up of small clusters into achiral monomers (rate $\lambda_0$), 
\item the break up of larger clusters into achiral monomers (rate $\lambda_1$). 
\end{itemize}

Such a model can exhibit symmetry-breaking to a solution in which 
$x_1\neq x_2$ and $x_2\neq y_2$.  Uwaha points out that the 
recycling part of the model (the $\lambda_*$ parameters) are crucial 
to the formation of a `completely' homochiral state. One problem with 
such a model is that since the variables are all total masses in the 
system, the size of clusters is not explicitly included.  In asymmetric 
distributions, the typical size of left- and right- handed clusters may 
differ drastically, hence the rates of reactions will proceed differently 
in the cases of a few large crystals or many smaller crystals. 

Sandars has proposed a model of symmetry-breaking in the 
formation of chiral polymers \cite{sandars}. His model has an 
achiral substrate ($S$) which splits into chiral monomers $L_1,R_1$ 
both spontaneously at a slow rate and at a faster rate, when catalysed 
by the presence of long homochiral chains.  This catalytic effect has 
both autocatalytic and crosscatalytic components, that is, for example, 
the presence of long right-handed chains $R_n$ autocatalyses the 
production of right-handed monomers $R_1$ from $S$, (autocatalysis) 
as well as the production of left-handed monomers, $L_1$ (crosscatalysis).   
Sandars assumes the growth rates of chains are linear and not 
catalysed; the other mechanism required to produce a symmetry-breaking 
bifurcation to a chiral state is cross-inhibition, by which chains of 
opposite handednesses interact and prevent either from further 
growth. These mechanisms are summarised by 
\beqa
S \rightarrow L_1 , && S \rightarrow R_1 , 
	\qquad \mbox{slow} , \nn \\ 
S \!+\! L_n \rightarrow L_1 \!+\! L_n , && 
S \!+\! R_n \rightarrow R_1 \!+\! R_n , \quad 
\mbox{autocatalytic, rate $\propto 1\!+\!f$} , \nn 
\\
S \!+\! R_n \rightarrow L_1 \!+\! R_n , && 
S \!+\! L_n \rightarrow R_1 \!+\! L_n , \quad 
\mbox{cross-catalytic, rate $\propto 1\!-\!f$} , \nn 
\\
L_n + L_1 \rightarrow L_{n+1} , && 
R_n + R_1 \rightarrow R_{n+1} , \qquad 
\mbox{chain growth, rate $=a$} , \nn
\\
L_n + R_1 \rightarrow Q_{n+1} , && 
R_n + L_1 \rightarrow P_{n+1} , \qquad 
\mbox{cross-inhibition, rate $=a\chi$} . \nn
\eeqa
This model and generalisations of it have been analysed 
by Sandars \cite{sandars},  Brandenburg \etal\  \cite{axel,axel3}, 
Multimaki \& Brandenburg \cite{axel2}, 
Wattis \& Coveney \cite{jw-sandars,jw-ch-rna-rev}, 
Gleiser \& Walker \cite{sara-bup}, Gleiser \etal\  \cite{sara-punc}. 
Typically a classic pitchfork bifurcation is found when the fidelity 
($f$) of the autocatalysis over the cross-catalysis is increased.  
One counterintuitive effect is that increasing the cross-inhibition 
effect ($\chi$) aids the bifurcation, allowing it to occur at lower 
values of the fidelity parameter $f$.

%------------------------------------------
\subsection{Experimental results on homochiralisation}

The Soai reaction was one of the first experiments which 
demonstrated that a chemical reaction could amplify initial small 
imbalances in chiral balance;  that is, a small enantiomeric exess in 
catalyst at the start of the experiment led to a much larger imbalance 
in the chiralities of the products at the end of the reaction. Soai \etal\ 
\cite{soai} was able to achieve an enantiomeric exess exceeding 
85\% in the asymmetric autocatalysis of chiral pyrimidyl alkanol.  

The first work showing that crystallisation experiments could exhibit 
symmetry breaking was that of Kondepudi \& Nelson 
\cite{kon-sci}.  Later Kondepudi \etal\ \cite{kon-jacs} 
showed that the stirring rate was a good bifurcation parameter to 
analyse the final distribution of chiralities of crystals emerging from a 
supersaturated solution of sodium chlorate.  With no stirring, there 
were approximately equal numbers of left- and right-handed crystals.  
Above a critical (threshold) stirring rate, the imbalance in the numbers 
of each handedness increased, until, at large enough stirring rates, 
total chiral purity was achieved.  This is due to all crystals in the 
system being derived from the same `mother' crystal, which is the first 
crystal to become  established in the system; all other crystals grow 
from fragments removed from it (either directly or indirectly). 
Before this, Kondepudi \& Nelson \cite{kon-pla,kon-nat}  
worked on the theory of chiral symmetry-breaking mechanisms 
with the aim of predicting how parity-violating perturbations could 
be amplified to give an enantiomeric exess in prebiotic chemistry, 
and the timescales involved.  Their results suggest 
a timescale of approximately $10^4$ years.   More recently, 
Kondepudi and Asakura \cite{kon+a} have summarised 
both the experimental and theoretical aspects of this work. 
% 1 TIDY UP KONDEPUDI REFERENCES %%

% 2 MORE DETAILS OF VIEDMA's EXPTS %% 
Viedma \cite{viedma} was the first to observe that grinding a 
mixture of chiral crystals eventually led to a distribution of crystals 
which were all of the same handedness.  The crystalline material 
used was sodium chlorate, as used by Kondepudi \etal\ 
\cite{kon-sci}.  
Samples of L and D crystals are mixed with water in round-bottomed 
flasks and the system is stirred by a magnetic bar (of length 3-20mm) 
at 600rpm.  The system is maintained in a supersaturated state; 
small glass balls are added to continually crush the crystals.  
The grinding is thus continuous, and crystals are maintained below 
a size of 200 $\mu$m. The chirality of the resulting crystals was 
determined by removing them from the flask, allowing them to grow 
and measuring their optical activity. 
The results show that, over time, the percentages of left- and 
right-handed crystals steadily change from about 50/50 to 100/0 or 
0/100  -- a state which is described as complete chiral purity. 
With stirring only and no glass balls, the systems conserve their 
initial chiral excesses; with glass balls present and stirring, the chiral
excess increases, and this occurs more rapidly if more balls are 
present or the speed of stirring is increased. 

More recently, Noorduin \etal\ \cite{wim} have observed a 
similar effect with amino acids -- a much more relevant molecule 
in the study of origins of life.  This work has been reviewed by 
McBride \& Tully \cite{mcbride-nature}, who add to the 
speculation on the mechanisms responsible for the phenomenon.  
Noorduin \etal\ describe grinding as `dynamic 
dissolution/crystallization processes that result in the conversion 
of one solid enantiomorph into the other'.  They also note that `once 
a state of single chirality is achieved, the system is ``locked'' 
because primary nucleation to form and sustain new crystals from 
the opposite enantiomer is kinetically prohibited'.  Both these quotes 
include the crucial fact that the process evolves {\em not} towards 
an equilibrium solution (which would be racemic), but towards a 
different, dynamic steady-state solution. As noted by Plasson 
(personal communication, 2008), this nonequilibrium state is 
maintained due to the constant input of energy into the system 
through the grinding process. 

McBride \& Tully \cite{mcbride-nature} discuss  the growth 
of one enantiomorph, and the dissolution of the other as a type of 
Ostwald ripening process; with the large surface area to volume ratio 
of smaller crystals giving a rapid dissolution rate, whilst larger crystals, 
have a lower surface area to volume ratio meaning that they dissolve 
more slowly. However appealing such an argument maybe, since 
surface area arguments can equally well be applied to the growth 
side of the process, it is not clear that this is either necessary or 
sufficient.   Infact, the model analysed later in this paper will show 
that a critical cluster size is not necessary to explain homochiralisation 
through grinding.  

%---------------------------------------------------------
\subsection{Our aims}

We aim to describe the results of the crystal grinding phenomenon 
through a model which recycles mass through grinding, which 
causes crystals to fragment, rather than having explicit mass input 
and removal.  Simultaneously we need crystal growth processes 
to maintain a distribution of sizeable crystals.  

% 3 MORE DETAIL ON HOW GRINDING FITS INTO ORIG OF LIFE %% 
We assume that the crystals are solids formed in an aqueous 
environment, however, we leave open questions as to whether 
they are crystals of some mineral of {\em direct} biological 
relevance (such as amino acids), or whether they are some other 
material, which after growing, will later provide a chirally 
selective surface for biomolecules to crystallise on, or be a catalyst 
for chiral polymerisation to occur.  Following Darwin's 
\cite{darwin} ``warm little pond'', an attractive scenario might 
be a tidal rock pool, where waves agitating pebbles provide the 
energetic input for grinding.  Taking more account of recent work, 
a more likely place is a suboceanic hydrothermal vent where 
the rapid convection of hot water impels growing nucleii into the 
vent's rough walls as well as breaking particles off the walls and 
entraining them into the fluid flow, simultaneously grinding any 
growing crystals. 

In Section \ref{model-sec} we propose a detailed microscopic 
model of the nucleation and crystal growth of several species 
simultaneously.  This has the form of a generalised Becker-D\"{o}ring 
system of equations \cite{bd}.  Due to the complexity of the 
model we immediately simplify it, making assumptions on the rate 
coefficients.   Furthermore, to elucidate those processes which 
are responsible for homochiralisation, we remove some processes 
completely so as to obtain a simple system of ordinary differential 
equations which can be analysed theoretically.  

The simplest model which might be expected to show 
homochiralisation is one which has small and large clusters of each 
handedness.  Such a truncated model is considered in Section 
\ref{tetra-sec} wherein it is shown that such a model might lead 
to amplification of enantiomeric exess in the short time, but that 
in the long-time limit, only the racemic state can be approached. 
This model has the structure akin to that of Saito \& Hyuga 
\cite{saito2} truncated at the tetramer level. 

Hence, in Section \ref{hex-sec} we consider a more complex model 
with a cut-off at larger sizes (one can think of small, medium, and large 
clusters of each handedness).  Such a model has a similar structure to 
the hexamer truncation analysed by Saito \& Hyuga \cite{saito2}.  
We find that such a model does allow a final steady-state in which one 
chirality dominates the system and the other is present only in 
vanishingly small amounts. 

However, as discussed earlier, there may be subtle effects whereby it 
is not just the {\em number} of crystals of each type that is important 
to the effect, but a combination of size and number of each handedness 
of crystal that is important to the evolution of the process.  Hence, in 
Section \ref{new-sec} we introduce an alternative reduction of the 
system of governing equations.  In this, instead of truncating and 
keeping only clusters of a small size, we postulate a form for the 
distribution which includes information on both the number and size 
of crystals, and use these two quantities to construct a system of 
five ordinary differential equations for the system's evolution. 

We discuss the results in Sections \ref{disc-sec} and \ref{conc-sec} 
which conclude the paper.  The Appendix \ref{app} shows how, 
by removing the symmetry in the growth rates of the two 
handednesses, the model could be generalised to account for 
the competitive nucleation of different polymorphs growing from 
a common supply of monomer. 

%---------------------------------------------------------
\section{The BD model with dimer interactions and 
an amorphous metastable phase} 
\label{model-sec}
\setcounter{equation}{0}

%--------------------------------------
\subsection{Preliminaries}

Smoluchowski \cite{smol} proposed a model in which clusters 
of any sizes could combine pairwise to form larger clusters. Chemically 
this process is written $C_r + C_s \rightarrow C_{r+s}$ where $C_r$ 
represents a cluster of size $r$.  Assuming this process is reversible 
and occurs with a forward rate given by $a_{r,s}$ and a reverse rate 
given by $b_{r,s}$, the law of mass action yields the kinetic equations 
\beqa
\frac{\dd c_r}{\dd t} &\!=\!&\! \half \sum_{s=1}^{r-1} 
 \left( a_{s,r-s} c_s c_{r-s} - b_{s,r-s} c_r \right) 
 - \sum_{s=1}^\infty \left( a_{r,s} c_r c_s - 
 b_{r,s} c_{r+s} \right) . \nn \\ && \lbl{smol-eq}
\eeqa
These are known as the coagulation-fragmentation equations.  
There are simplifications in which only interactions between clusters 
of particular sizes are permitted to occur, for example when only 
cluster-monomer interactions can occur, the Becker-D\"{o}ring 
equations \cite{bd} are obtained.  Da Costa has formulated a 
system in which only clusters upto a certain size ($N$) are permitted 
to coalesce with or fragment from other clusters.  In the case of 
$N=2$, which is pertinent to the current study, only cluster-monomer 
and cluster-dimer interactions are allowed, for example 
\beq
C_r + C_1 \rightleftharpoons C_{r+1} , 
\qquad 
C_r + C_2 \rightleftharpoons C_{r+2} . 
\eeq  
This leads to a system of kinetic equations of the form 
\beqa
\frac{\dd c_r}{\dd t} & = & J_{r-1} - J_r + K_{r-2} - K_r , 
	\qquad (r\geq3) , \lbl{gbd-eq1} \\ 
\frac{\dd c_2}{\dd t} & = & J_1 - J_2 - K_2 
	- \ds\sum_{r=1}^\infty K_r , \\ 
\frac{\dd c_1}{\dd t} & = & - J_1 - K_2 
	- \ds\sum_{r=1}^\infty J_r , \\ 
J_r &=& a_r c_r c_1 - b_{r+1} c_{r+1} , \qquad 
K_r = \alpha_r c_r c_2 - \beta_{r+2} c_{r+2} . 
\lbl{gbd-eq4} \eeqa
A simple example of such a system has been analysed 
previously by Bolton \& Wattis \cite{bw-dimers}. 

In the next subsection we generalise the model (\ref{smol-eq}) to 
include a variety of `species' or `morphologies' of cluster, 
representing left-handed, right-handed and achiral clusters.  We 
simplify the model in stages to one in which only monomer and dimer 
interactions are described, and then one in which only dimer 
interactions occur. 

%----------------------------------------------------
\subsection{A full microscopic model of chiral crystallisation}

We start by outlining all the possible cluster growth, fragmentation 
and transformation processes.  We denote the two handed clusters 
by $X_r$, $Y_r$, where the subscript $r$ specifies the size of cluster. 
Achiral clusters are denoted by $C_r$, and we allow clusters to 
change their morphology spontaneously according to 
\beq \begin{array}{rclclccrclcl}
C_r & \rightarrow & X_r & \quad& {\rm rate} = \mu_r , &&
X_r & \rightarrow & C_r & \quad& {\rm rate} = \mu_r \nu_r , \\ 
C_r & \rightarrow & Y_r & \quad& {\rm rate} = \mu_r , && 
Y_r & \rightarrow & C_r & \quad& {\rm rate} = \mu_r \nu_r .  
\end{array} \eeq
We allow clusters to grow by coalescing with clusters 
of similar handedness or an achiral cluster.  In the case of 
the latter process, we assume that the cluster produced 
is chiral with the same chirality as the parent. Thus 
\beq \begin{array}{rclcl}
X_r + X_s & \rightarrow & X_{r+s} , && {\rm rate} = \xi_{r,s}, \\ 
X_r + C_s & \rightarrow & X_{r+s} , && {\rm rate} = \alpha_{r,s},\\ 
C_r + C_s & \rightarrow & C_{r+s} , && {\rm rate} = \delta_{r,s},\\ 
Y_r + C_s & \rightarrow & Y_{r+s} , && {\rm rate} = \alpha_{r,s},\\ 
Y_r + Y_s & \rightarrow & Y_{r+s} , && {\rm rate} = \xi_{r,s} . 
\end{array} \eeq
We do not permit clusters of opposite to chirality to merge. 
Finally we describe fragmentation: all clusters may 
fragment, producing two smaller clusters each of 
the same chirality as the parent cluster 
\beq \begin{array}{rclcl}
X_{r+s} & \rightarrow & X_r + X_s && {\rm rate} = \beta_{r,s}, \\ 
C_{r+s} & \rightarrow & C_r + C_s && {\rm rate} = \epsilon_{r,s}, \\ 
Y_{r+s} & \rightarrow & Y_r + Y_s &\quad& {\rm rate} = \beta_{r,s} . 
\end{array} \eeq
Setting up concentration variables for each size and each type of 
cluster by defining $c_r(t) = [C_r]$, $x_r(t) = [X_r]$, $y_r(t) = [Y_r]$ 
and applying the law of mass action, we obtain 
\beqa
\frac{\dd c_r}{\dd t} &\!=\!& -2\mu_r c_r + \mu_r\nu_r(x_r+y_r) 
	- \sum_{k=1}^\infty \alpha_{k,r} c_r (x_k+y_k) \\ && \nn 
	+ \half \sum_{k=1}^{r-1} \left( \delta_{k,r-k} c_k c_{r-k} 
	- \epsilon_{k,r-k} c_k c_{r-k} \right) 
	- \sum_{k=1}^\infty \left( \delta_{k,r} c_k c_r 
	- \epsilon_{k,r} c_{r+k} \right) , 
\lbl{gbd1} \\ 
\frac{\dd x_r}{\dd t} &\!=\!&\! \mu_r c_r \!-\! \mu_r \nu_r x_r 
	+ \sum_{k=1}^{r-1} \alpha_{k,r-k} c_k x_{r-k} 
	\!-\! \half \sum_{k=1}^{r-1} \left( \xi_{k,r-k} x_k x_{r-k} 
	\!-\! \beta_{k,r\!-\!k} x_r \right) \nn \\ && 
	- \sum_{k=1}^\infty \left( \xi_{k,r} x_k x_r 
	- \beta_{k,r} x_{r+k} \right) , 
\\ 
\frac{\dd y_r}{\dd t} &\!=\!&\! \mu_r c_r \!-\! \mu_r \nu_r y_r  
	+ \sum_{k=1}^{r-1} \alpha_{k,r-k} c_k y_{r-k} 
	\!-\! \half \sum_{k=1}^{r-1} \left( \xi_{k,r-k} y_k y_{r-k} 
	\!-\! \beta_{k,r\!-\!k} y_r \right) \nn \\ && 
	- \sum_{k=1}^\infty \left( \xi_{k,r} y_k y_r 
	- \beta_{k,r} y_{r+k} \right) . 
\lbl{gbd3} \eeqa
The main problem with such a model is the vast number 
of parameters that have been introduced ($\alpha_{r,k}$, 
$\xi_{r,k}$, $\beta_{r,k}$, $\mu_r$, $\nu_r$, $\delta_{r,k}$, 
$\epsilon_{r,k}$, for all $k,r$).  

Hence we make several simplifications: 
\begin{description} 
\item[(i)] 
we assume that the dominant coagulation and fragmentation 
processes are between large and very small clusters (rather 
than large clusters and other large clusters).   Specifically, we 
assume that only coalescences involving $C_1$ and $C_2$ 
need to be retained in the model, and fragmentation always 
yields either a monomer or a dimer fragment.  This assumption 
means that the system can be reduced to a generalised 
Becker-D\"{o}ring equation closer to the form of 
(\ref{gbd-eq1})--(\ref{gbd-eq4}) rather than (\ref{smol-eq}); 

\item[(ii)] 
we also assume that the achiral clusters are unstable at larger size, 
so that their presence is only relevant at small sizes.  
Typically at small sizes, clusters are amorphous and do not take 
on the properties of the bulk phase, hence at small sizes clusters 
can be considered achiral.   We assume that there is a regime of 
cluster sizes where there is a transition to chiral structures, and 
where clusters can take on the bulk structure (which is chiral) as well 
as exist in amorphous form.   At even larger sizes, we assume that 
only the chiral forms exist, and no achiral structure can be adopted;  

\item[(iv)] 
furthermore, we assume that all rates are independent of cluster size, 
specifically, 
\beqa
\alpha_{_{k,1}} & = & a , \qquad \qquad 
\alpha_{_{k,2}} = \alpha , \qquad \quad 
\alpha_{_{k,r}} =0 , \quad (r\geq2) 	\\ 
\mu_2 &=& \mu , \qquad \qquad 
\mu_r=0 , \quad (r\geq3) , 	\\ 
\nu_2 &=& \nu , \qquad \qquad 
\nu_r=0 , \quad (r\geq3) , 	\\ 
\delta_{1,1} & = & \delta , \qquad 
\delta_{k,r} = 0 , \quad ({\rm otherwise}) 	\\ 
\epsilon_{1,1} & = & \epsilon , \qquad 
\epsilon_{k,r} = 0 , \quad ({\rm  otherwise})	\\ 
\xi_{k,2} &=& \xi_{2,k} = \xi , \qquad 
\xi_{k,r} = 0 , \quad ({\rm otherwise})		\\
\beta_{k,1} & = & \beta_{1,k} = b , \qquad 
\beta_{k,2} = \beta_{2,k} = \beta , \qquad 
\beta_{k,r} = 0 , \quad ({\rm otherwise}), 
\nn \\ && \eeqa
Ultimately we will set $a=b=0=\delta=\epsilon$ so that we have only 
five parameters to consider ($\alpha$, $\xi$, $\beta$, $\mu$, $\nu$). 

\end{description} 

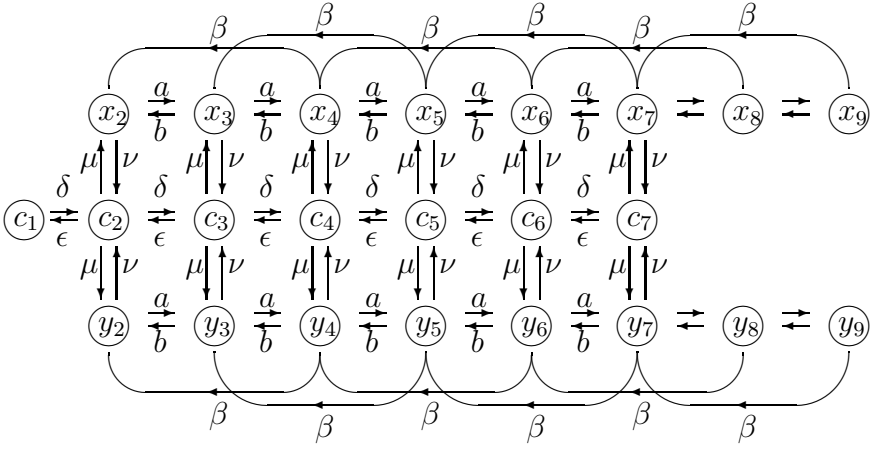
\begin{figure}[!ht]
\begin{picture}(500,160)(35,-35)
\put(48,50){\circle{15}}
\multiput(80,50)(40,0){6}{\circle{15}}
\multiput(80,10)(40,0){8}{\circle{15}}
\multiput(80,90)(40,0){8}{\circle{15}}
\put(044,48){$c_1$}
\put(076,48){$c_2$}
\put(116,48){$c_3$}
\put(156,48){$c_4$}
\put(196,48){$c_5$}
\put(236,48){$c_6$}
\put(276,48){$c_7$}
\put(076,88){$x_2$}
\put(116,88){$x_3$}
\put(156,88){$x_4$}
\put(196,88){$x_5$}
\put(236,88){$x_6$}
\put(276,88){$x_7$}
\put(316,88){$x_8$}
\put(356,88){$x_9$}
\put(076,08){$y_2$}
\put(116,08){$y_3$}
\put(156,08){$y_4$}
\put(196,08){$y_5$}
\put(236,08){$y_6$}
\put(276,08){$y_7$}
\put(316,08){$y_8$}
\put(356,08){$y_9$}
\multiput(83,20)(40,0){6}{\vector(0,1){20}}
\multiput(77,40)(40,0){6}{\vector(0,-1){20}}
\multiput(77,60)(40,0){6}{\vector(0,1){20}}
\multiput(83,80)(40,0){6}{\vector(0,-1){20}}
\multiput(69,30)(40,0){6}{$\mu$}
\multiput(85,30)(40,0){6}{$\nu$}
\multiput(69,70)(40,0){6}{$\mu$}
\multiput(85,70)(40,0){6}{$\nu$}
\put(058,53){\vector(1,0){10}}
\put(068,50){\vector(-1,0){10}}
\multiput(095,53)(40,0){5}{\vector(1,0){10}}
\multiput(105,50)(40,0){5}{\vector(-1,0){10}}
\multiput(095,95)(40,0){7}{\vector(1,0){10}}
\multiput(105,90)(40,0){7}{\vector(-1,0){10}}
\multiput(095,15)(40,0){7}{\vector(1,0){10}}
\multiput(105,10)(40,0){7}{\vector(-1,0){10}}
\put(060,59){$\delta$}
\put(060,40){$\epsilon$}
\multiput(097,59)(40,0){5}{$\delta$}
\multiput(097,40)(40,0){5}{$\epsilon$}
\multiput(097,97)(40,0){5}{$a$}
\multiput(097,80)(40,0){5}{$b$}
\multiput(097,17)(40,0){5}{$a$}
\multiput(097,00)(40,0){5}{$b$}
\put(118,119){$\beta$}
\put(127,115){\vector(-1,0){10}}
\put(120,100){\oval(80,30)[t]}
\put(120,  0){\oval(80,30)[b]}
\put(127,-15){\vector(-1,0){10}}
\put(118,-27){$\beta$}
\put(158,124){$\beta$}
\put(167,120){\vector(-1,0){10}}
\put(160,100){\oval(80,40)[t]}
\put(160,  0){\oval(80,40)[b]}
\put(167,-20){\vector(-1,0){10}}
\put(158,-32){$\beta$}
\put(198,119){$\beta$}
\put(207,115){\vector(-1,0){10}}
\put(200,100){\oval(80,30)[t]}
\put(200,  0){\oval(80,30)[b]}
\put(207,-15){\vector(-1,0){10}}
\put(198,-27){$\beta$}
\put(238,124){$\beta$}
\put(247,120){\vector(-1,0){10}}
\put(240,100){\oval(80,40)[t]}
\put(240,  0){\oval(80,40)[b]}
\put(247,-20){\vector(-1,0){10}}
\put(238,-32){$\beta$}
\put(278,119){$\beta$}
\put(287,115){\vector(-1,0){10}}
\put(280,100){\oval(80,30)[t]}
\put(280,  0){\oval(80,30)[b]}
\put(287,-15){\vector(-1,0){10}}
\put(278,-27){$\beta$}
\put(318,124){$\beta$}
\put(327,120){\vector(-1,0){10}}
\put(320,100){\oval(80,40)[t]}
\put(320,  0){\oval(80,40)[b]}
\put(327,-20){\vector(-1,0){10}}
\put(318,-32){$\beta$}
\end{picture} 
\caption{Reaction scheme involving monomer and 
dimer aggregation and fragmentation of achiral clusters 
and those of both handednesses (right and left). 
The aggregation of achiral and chiral clusters 
is not shown (rates $\alpha$, $\xi$). }
\label{rec-sch-fig}
\end{figure}

This scheme is illustrated in Figure \ref{rec-sch-fig}.  However, 
before writing down a further system of equations,  we make one 
further simplification. We take the transition region described in (ii), 
above, to be just the dimers.  Thus the only types of achiral cluster 
are the monomer and the dimer ($c_1$, $c_2$); dimers exist in achiral, 
right- and left-handed forms ($c_2$, $x_2$, $y_2$); at larger sizes 
only left- and right-handed clusters exist ($x_r$, $y_r$, $r\geq2$). 

The kinetic equations can be reduced to 
\beqa 
\frac{\dd c_1}{\dd t} & = & 2 \ep c_2 - 2 \de c_1^2 
	- \sum_{r=2}^\infty ( a c_1 x_r + a c_1 y_r - b x_{r+1} 
	- b y_{r+1} ) , \lbl{gbd-c1}
	\\ 
\frac{\dd c_2}{\dd t} & = & \de c_1^2 - \ep c_2 - 2 \mu c_2 
	+ \mu\nu (x_2+y_2) - \sum_{r=2}^\infty \alpha c_2 (x_r+y_r) , 
	\\
\frac{\dd x_r}{\dd t} & = & a c_1 x_{r-1} - b x_r 
	- a c_1 x_r + b x_{r+1} + \alpha c_2 x_{r-2} 
	- \alpha c_2 x_r  \nn\\ && 
	- \beta x_r + \beta x_{r+2} + \xi x_2 x_{r-2}  
	- \xi x_2 x_r , \qquad \hfill (r\geq4) , 
	\\ 
\frac{\dd x_3}{\dd t} & = & a c_1 x_2 - b x_3 - a c_1 x_3 
	+ b x_4 - \alpha c_2 x_3 - \xi x_2 x_3 + \beta x_5 , 
	\\ 
\frac{\dd x_2}{\dd t} & = & \mu c_2 - \mu\nu x_2 + b x_3 
	- a c_1 x_2 - \alpha x_2 c_2 
	+ \beta x_4 \nn \\ && + \sum_{r=2}^\infty \beta x_{r+2} 
	- \sum_{r=2}^\infty \xi x_2 x_r - \xi x_2^2 , 
	\\ 
\frac{\dd y_r}{\dd t} & = & a c_1 y_{r-1} - b y_r 
	- a c_1 y_r + b y_{r+1} + \alpha c_2 y_{r-2} 
	- \alpha c_2 y_r   \nn\\ && 
	- \beta y_r + \beta y_{r+2} + \xi y_2 y_{r-2} 
	- \xi y_2 y_r , \qquad \hfill (r\geq4), 
	\\ 
\frac{\dd y_3}{\dd t} & = & a c_1 y_2 - b y_3 - a c_1 y_3 
	+ b y_4 - \alpha c_2 y_3 - \xi y_2 y_3 + \beta y_5  , 
	\\ 
\frac{\dd y_2}{\dd t} & = & \mu c_2 - \mu\nu y_2 + b y_3 
	- a c_1 y_2 - \alpha y_2 c_2 
	+ \beta y_4 \nn \\ && + \sum_{r=2}^\infty \beta y_{r+2} 
	- \sum_{r=2}^\infty \xi y_2 y_r - \xi y_2^2 . \lbl{gbd-y2}
\eeqa 

%--------------------------------------------------------------------
\subsection{Summary and simulations of the macroscopic model}
\label{macro-sec} 

The advantage of the above simplifications is that certain sums 
appear repeatedly; by defining new quantities as these sums, 
the system can be written in a simpler fashion. We define $N_x = 
\sum_{r=2}^\infty x_r$, $N_y = \sum_{r=2}^\infty y_r$, then 
\beqa
\frac{\dd c_1}{\dd t} & = & 2 \ep c_2 - 2 \de c_1^2 
	- a c_1 (N_x+N_y) + b (N_x-x_2+N_y-y_2) , \lbl{macro-c1}\\ 
\frac{\dd c_2}{\dd t} & = & \de c_1^2 - \ep c_2 - 2 \mu c_2 
	+ \mu\nu (x_2+y_2) - \alpha c_2(N_x+N_y) ,\\
\frac{\dd N_x}{\dd t} & = & \mu c_2 - \mu\nu x_2 
	+ \beta (N_x-x_3-x_2) - \xi x_2 N_x , \\ 
\frac{\dd x_2}{\dd t} & = & \mu c_2 - \mu\nu x_2 + b x_3 
	- a c_1 x_2 - \alpha x_2 c_2 + \beta (x_4+N_x-x_2-x_3) 
	\nn \\ && -\xi x_2^2 - \xi x_2 N_x , \\ 
\frac{\dd N_y}{\dd t} & = & \mu c_2 - \mu\nu y_2 
	+ \beta (N_y-y_3-y_2) - \xi y_2 N_y , \\ 
\frac{\dd y_2}{\dd t} & = & \mu c_2 - \mu\nu y_2 + b y_3 
	- a c_1 y_2 - \alpha y_2 c_2 + \beta (y_4+N_y-y_2-y_3) 
	\nn\\ && - \xi y_2^2 - \xi y_2 N_y . \lbl{macro-y2}
\eeqa 
However, such a system of equations is not `closed'.  The equations 
contain $x_3,y_3,x_4,y_4$, and yet we have no expressions for 
these; reintroducing equations for $x_3,y_3$ would introduce 
$x_5,y_5$ and so an infinite regression would be entered into.  

\begin{figure}[!ht]
\vspace*{65mm}
\includegraphics{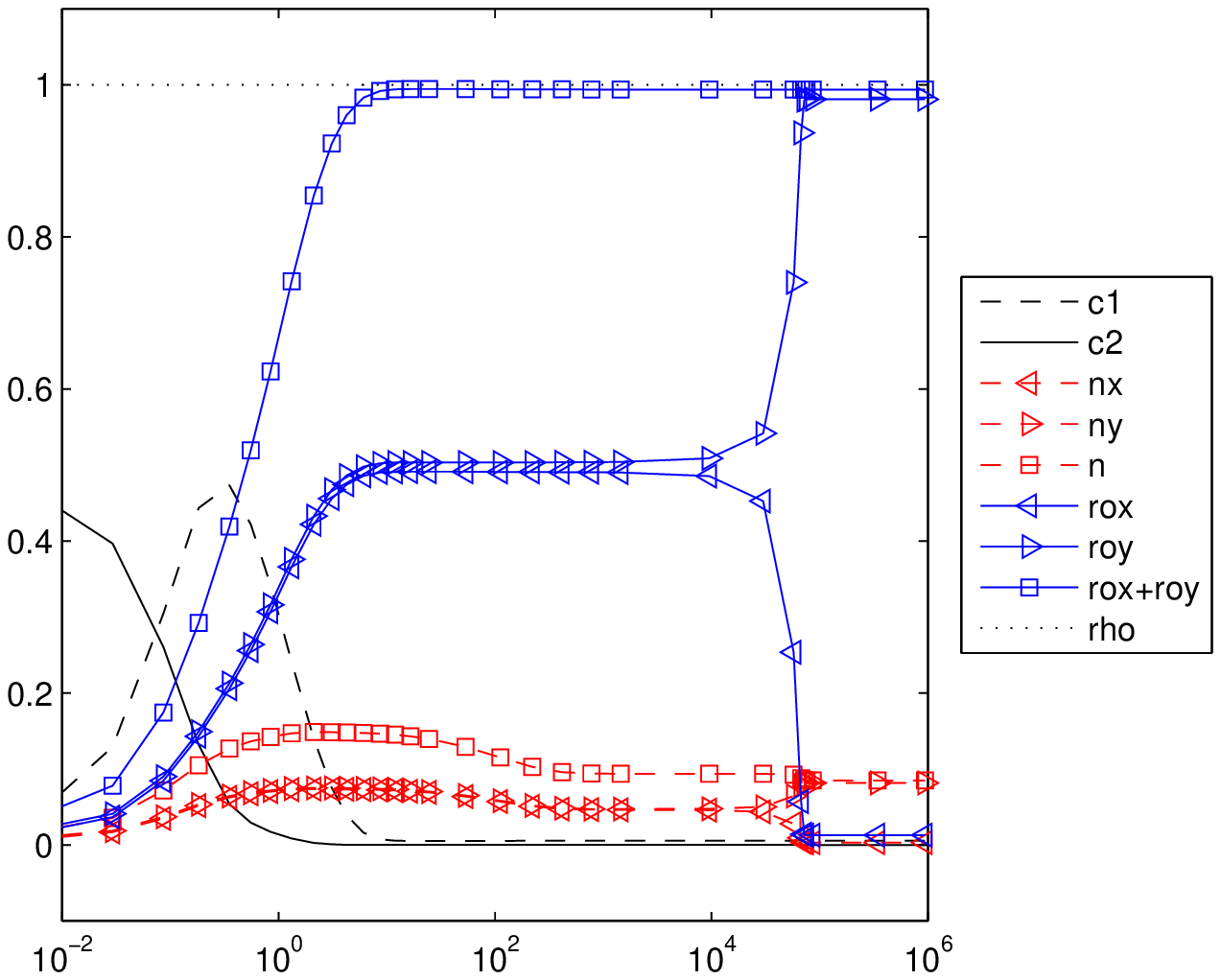}
\caption{ Plot of the concentrations $c_1$, $c_2$, 
$N_x$, $N_y$, $N=N_x+N_y$, $\ro_x$, $\ro_y$, 
$\ro_x+\ro_y$ and $\ro_x+\ro_y+2c_2+c1$ 
against time, $t$ on a logarithmic timescale.  Since model equations 
are in nondimensional form, the time units are arbitrary.  
Parameter values $\mu=1.0$, $\nu=0.5$, $\delta=1$, $\ep=5$, 
$a=4$, $b=0.02$, $\alpha=10$, $\xi=10$, $\beta=0.03$, with 
initial conditions $c_2=0.49$, $x_4(0)=0.004$, $y_4(0)=0.006$, 
and all other concentrations zero.  }
\label{fig-fullt}
\end{figure}

\begin{figure}[!ht]
\vspace*{73mm}
\includegraphics{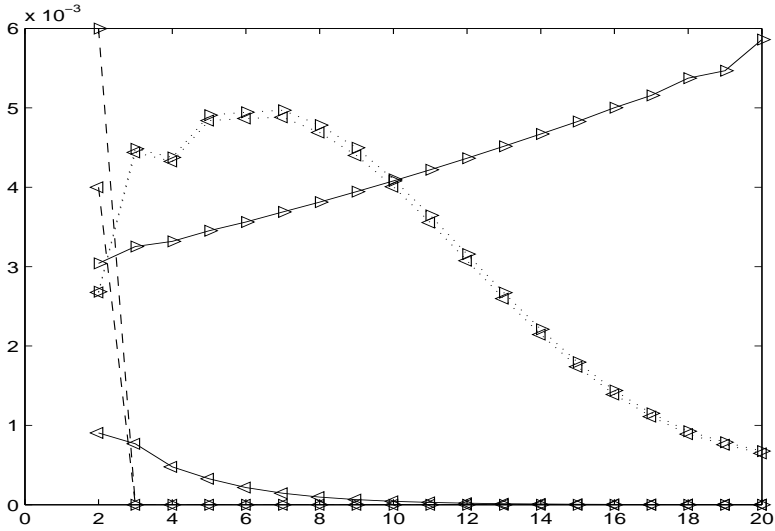}
\caption{Plot of the cluster size distribution at 
$t=0$ (dashed line), $t=112$ (dotted line) and 
$t=9.4\times10^5$. Parameters and initial conditions 
as in Figure \protect\ref{fig-fullt}. }
\label{fig-fullxy}
\end{figure}

Hence we need to find some suitable alternative expressions for 
$x_3,y_3,x_4,y_4$; or an alternative way of reducing the system to 
just a few ordinary differential equations that can easily be analysed. 
Such systems are considered in Sections \ref{tetra-sec}, 
\ref{hex-sec} and \ref{new-sec}. Before that, however, we illustrate 
the behaviour of the system by briefly presenting the results of some 
numerical simulations.    In Figures \ref{fig-fullt} and \ref{fig-fullxy} 
we show the results of a simulation of (\ref{macro-c1})--(\ref{macro-y2}). 
The former shows the evolution of the concentrations $c_1$ which 
rises then decays, $c_2$ which decays since the parameters have 
been chosen to reflect a cluster-dominated system. Also plotted are 
the numbers of clusters $N_x,N_y$ and the mass of material in 
clusters $\ro_x$, $\ro_y$ defined by 
\beq 
\ro_x = \sum_{j=2}^K j x_j , \qquad \ro_y = \sum_{j=2}^K j y_j . 
\eeq 
Note that under this definition $\ro_x+\ro_y+c_1+2c_2$ 
is conserved, and this is plotted as {\sl rho}.   Both the total number 
of clusters, $N_x+N_y$, and total mass of material in handed 
clusters $\ro_x+\ro_y$ appear to equilibrate by $t=10^2$, 
however, at a much later time ($t\sim 10^4 - 10^5$) a 
symmetry-breaking bifurcation occurs, and the system changes from 
almost racemic (that is, symmetric) to asymmetric.   This is more 
clearly seen in Figure \ref{fig-fullxy}, where we plot the cluster size 
distribution at three time points.   At $t=0$ there are only dimers 
present (dashed line), and we impose a small difference in the 
concentrations of $x_2$ and $y_2$.  At a later time, $t=112$ 
(dotted line), there is almost no difference between the $X$- and 
$Y$-distributions, however by the end of the simulation ($t\sim10^6$, 
solid line) one distribution clearly completely dominates the other. 

%--------------------------------------------------------------------
\subsection{Simplified macroscopic model}

To obtain the simplest model which involves three polymorphs 
corresponding to right-handed and left-handed chiral clusters and 
achiral clusters, we now aim to simplify the processes of cluster 
aggregation and fragmentation in (\ref{macro-c1})--(\ref{macro-y2}).  
Our aim is to retain the symmetry-breaking phenomenon but 
eliminate physical processes which are not necessary for it to occur. 

Our first simplification is to remove all clusters of odd size from the 
model, and just consider dimers, tetramers, hexamers, {\em etc}. 
This corresponds to putting $a=0$, $b=0$ which removes $x_3$ 
and $y_3$ from the system. 
Furthermore, we put $\ep=0$ and make $\delta$ large, so that the 
achiral monomer is rapidly and irreversibly converted to achiral dimer.  
Since the monomers do not then influence the evolution of any of 
the other variables, we further simplify the system by ignoring $c_1$ 
(or, more simply, just impose initial data in which $c_1(0)=0$).  
Thus we are left with 
\beqa
\!\!\!\frac{\dd c_2}{\dd t} & \!=\! & - 2 \mu c_2 
	+ \mu\nu (x_2+y_2) - \alpha c_2(N_x+N_y) , \lbl{smm2} \\
\!\!\!\frac{\dd N_x}{\dd t} & \!=\! & \mu c_2 - \mu\nu x_2 
	+ \beta (N_x-x_2) - \xi x_2 N_x , \\ 
\!\!\!\frac{\dd x_2}{\dd t} & \!=\! & \!\mu c_2 - \mu\nu x_2 - \alpha x_2 c_2 
	+ \beta (N_x\!-\!x_2 \!+\! x_4 ) - \xi x_2^2 - \xi x_2 N_x , \\ 
\!\!\!\frac{\dd N_y}{\dd t} & \!=\! & \mu c_2 - \mu\nu y_2 
	+ \beta (N_y-y_2) - \xi y_2 N_y , \\ 
\!\!\!\frac{\dd y_2}{\dd t} & \!=\! & \!\mu c_2 - \mu\nu y_2  - \alpha y_2 c_2 
	+ \beta (N_y\!-\!y_2 \!+\! y_4) - \xi y_2^2 - \xi y_2 N_y . \lbl{smmy2}
\eeqa 
Since we have removed four parameters from the model, 
and halved the number of dependent variables, we show a couple 
of numerical simulations just to show that the system 
above does still exhibit symmetry-breaking behaviour. 

\begin{figure}[!ht]
\vspace*{69mm}
\includegraphics{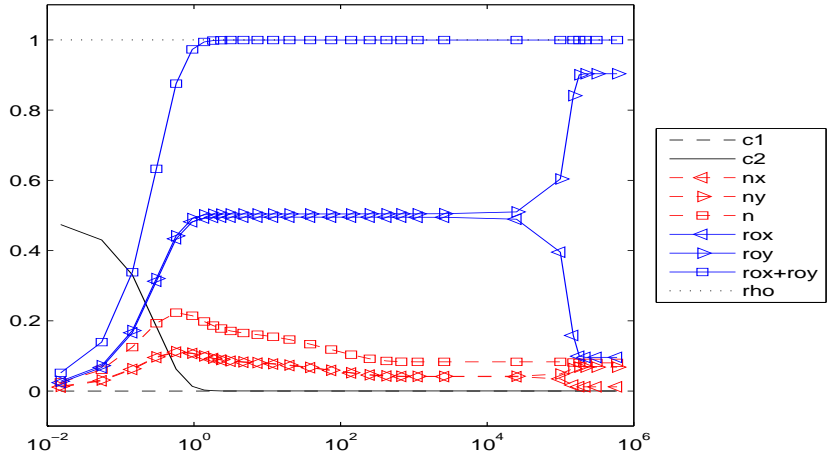}
\caption{ Plot of the concentrations $c_1$, $c_2$, 
$N_x$, $N_y$, $N=N_x+N_y$, $\ro_x$, $\ro_y$, 
$\ro_x+\ro_y$ and $\ro_x+\ro_y+2c_2+c_1$ 
against time, $t$ on a logarithmic timescale.  Since model 
equations are in nondimensional form, the time units are 
arbitrary.  Parameter values $\mu=1$, $\nu=0.5$,  
$\alpha=10$, $\xi=10$, $\beta=0.03$, with initial 
conditions $c_2=0.49$, $x_4(0)=0.004$, $y_4(0)=0.006$, 
all other concentrations zero.  }
\label{fig-dimt}
\end{figure}

\begin{figure}[!ht]
\vspace*{71mm}
\includegraphics{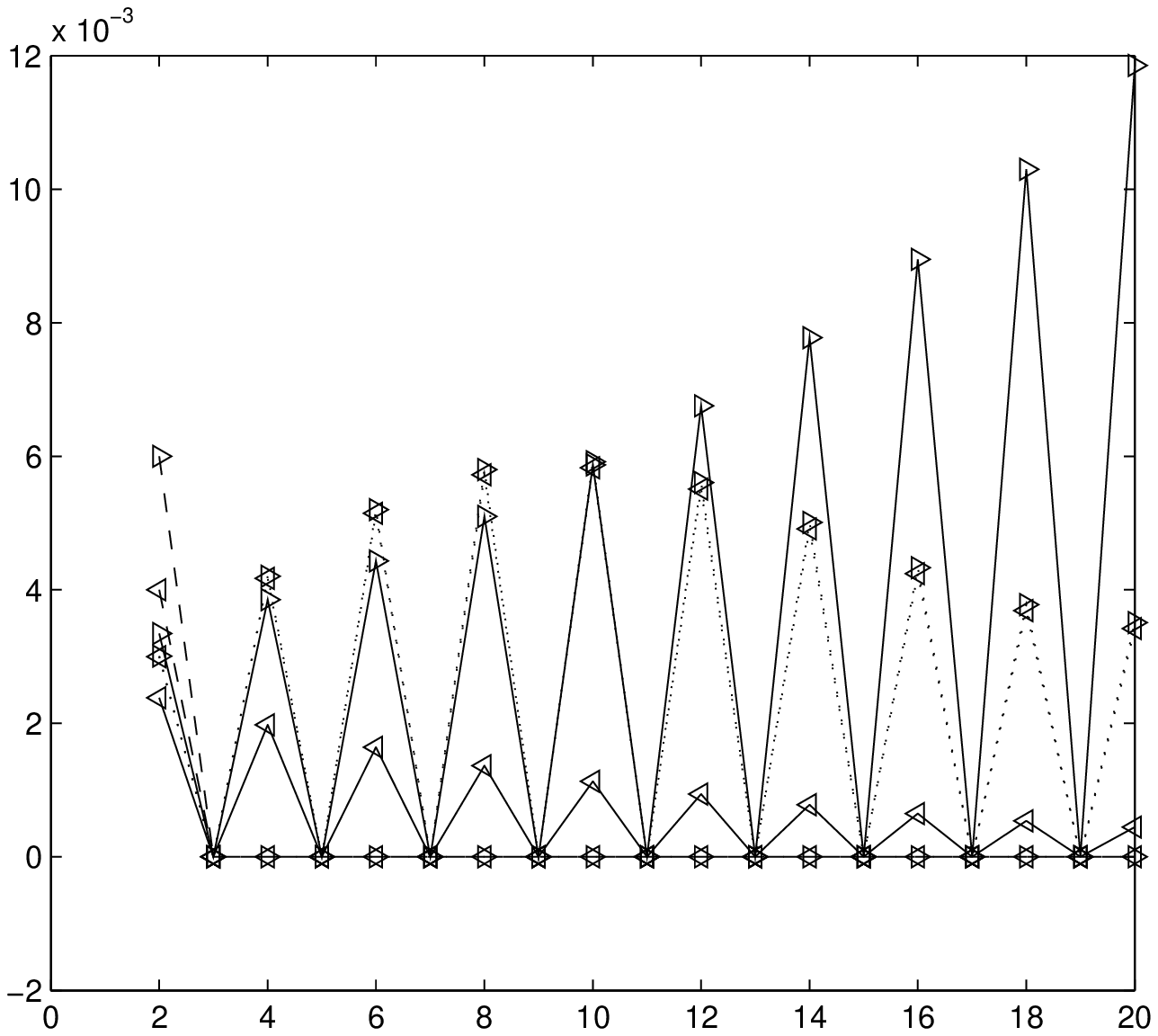}
\caption{Plot of the cluster size distribution at 
$t=0$ (dashed line), $t=250$ (dotted line) and  
$t=6\times10^5$. Parameters and initial conditions 
as in Figure \protect\ref{fig-dimt}. }
\label{fig-dimxy}
\end{figure}

\begin{figure}[!ht]
\vspace*{70mm}
\includegraphics{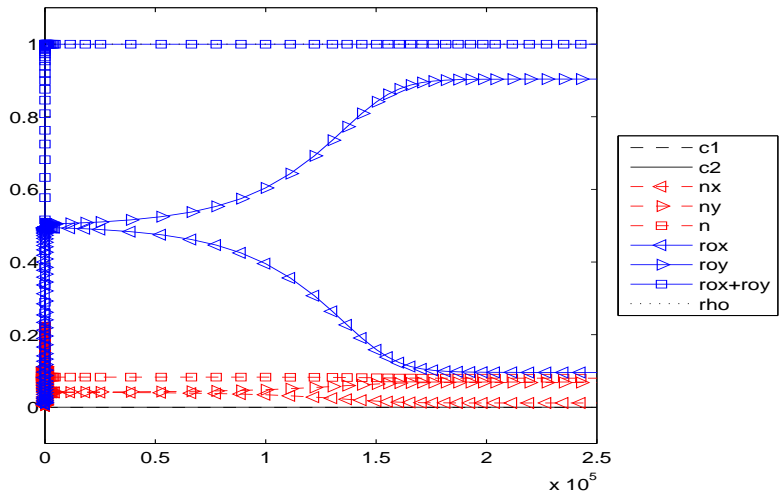}
\caption{ Plot of the concentrations $c_1$, $c_2$, $N_x$, $N_y$, 
$N=N_x+N_y$, $\ro_x$, $\ro_y$, $\ro_x+\ro_y$ and 
$\ro_x+\ro_y+2c_2+c_1$ against time, $t$ on a logarithmic timescale.  
Parameters and initial conditions as in Figure \protect\ref{fig-dimt}. }
\label{fig-dimtt}
\end{figure}

Figure \ref{fig-dimt} appears similar to Figure \ref{fig-fullt}, 
suggesting that removing the monomer interactions has changed the 
underlying dynamics little.  We still observe the characteristic 
equilibration of cluster numbers and cluster masses as $c_2$ decays, 
and then a period of quiesence ($t\sim10$ to $10^4$) before a later 
symmetry-breaking event, around $t\sim10^5$. 
At first sight, the distribution of $X$- and $Y$-clusters displayed in 
Figure \ref{fig-dimxy} is quite different to Figure \ref{fig-fullxy}; this 
is due to the absence of monomers from the system, meaning that 
only even-sized clusters can now be formed.  If one only looks at the 
even-sized clusters in Figure \ref{fig-dimxy}, we once again see only 
a slight difference at $t=0$ (dashed line), almost no difference at 
$t\approx250$ (dotted line) but a significant difference at 
$t=6\times10^5$ (solid line). 
We include one further graph here, Figure \ref{fig-dimtt} similar to 
Figure \ref{fig-dimt} but on a linear rather than a logarithmic timescale.  
This should be compared with Figures such as Figures 3 and 4 of 
Viedma \cite{viedma} and Figure 1 of Noorduin \etal\ \cite{wim}. 

%-----------------------------------------------------------------
\section{The truncation at tetramers} 
\label{tetra-sec} 
\setcounter{equation}{0}

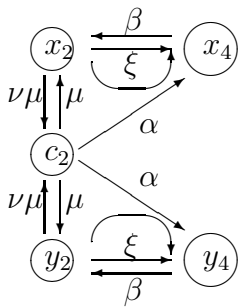
\begin{figure}[!ht]
\begin{picture}(500,120)(0,-10)
%
%\put(40,50){\circle{14}}
%\put(037,48){$c_1$}
\put(80,50){\circle{17}}
\put(077,48){$c_2$}
\put(080,10){\circle{17}}
\put(077,08){$y_2$}
\put(080,90){\circle{17}}
\put(077,88){$x_2$}
\put(140,10){\circle{20}}
\put(137,08){$y_4$}
\put(140,90){\circle{20}}
\put(137,88){$x_4$}
\put(77,20){\vector(0,1){20}}
\put(83,40){\vector(0,-1){20}}
\put(83,60){\vector(0,1){20}}
\put(77,80){\vector(0,-1){20}}
\put(63,30){$\nu\mu$}
\put(85,30){$\mu$}
\put(63,70){$\nu\mu$}
\put(85,70){$\mu$}
%\put(055,53){\vector(1,0){10}}
%\put(057,59){$p$}
%
\put(107,97){$\beta$}
\put(125,95){\vector(-1,0){30}}
\put(095,90){\vector(1,0){30}}
\put(107,81){$\xi$}
\put(110,85){\oval(30,20)[b]}
\put(125,85){\vector(0,1){5}}
\put(113,58){$\alpha$}
\put(090,52){\vector(3,+2){40}}
\put(090,48){\vector(3,-2){40}}
\put(113,38){$\alpha$}
\put(125,17){\vector(0,-1){5}}
\put(110,17){\oval(30,20)[t]}
\put(107,12){$\xi$}
\put(095,10){\vector(1,0){30}}
\put(125,05){\vector(-1,0){30}}
\put(107,-5){$\beta$}
\end{picture} 
\caption{Simplest possible reaction scheme which might 
exhibit chiral symmetry-breaking. }
\label{simp-rec-sch-fig}
\end{figure}

The simplest possible reaction scheme of the form 
(\ref{gbd-c1})--(\ref{gbd-y2}) which we might expect to exhibit 
symmetry-breaking to homochirality is the system truncated at 
tetramers, namely 
\beqa 
\ds\frac{\dd c_2}{\dd t} & = & - 2\mu c_2 + \mu\nu (x_2+y_2) 
	-\alpha c_2(x_2+y_2) , \lbl{dim-c2dot} \\ 
\ds\frac{\dd x_2}{\dd t} & = & \mu c_2 - \mu\nu x_2 
	- \alpha c_2 x_2 - 2 \xi x_2^2 + 2 \beta x_4 , \\ 
\ds\frac{\dd y_2}{\dd t} & = & \mu c_2 - \mu\nu y_2 
	- \alpha c_2 y_2 - 2 \xi y_2^2 + 2 \beta y_4 , \\ 
\ds\frac{\dd x_4}{\dd t} & = & \alpha x_2 c_2 + \xi x_2^2 - \beta x_4 , \\ 
\ds\frac{\dd y_4}{\dd t} & = & \alpha y_2 c_2 + \xi y_2^2 - \beta y_4 . 
\lbl{dim-y4dot}  \eeqa 

We investigate the symmetry-breaking by transforming 
the variables $x_2$, $x_4$, $y_2$, $y_4$ according to 
\beqa
x_2 = \half z (1+\theta) , &\quad& y_2 = \half z (1-\theta) , 
\lbl{tet-ztheta} \\  
x_4 = \half w (1+\phi) , && y_4 = \half w (1-\phi) , 
\lbl{tet-wphi} 
\eeqa
where $z=x_2+y_2$ is the total concentration of chiral dimers, 
$w=x_4+y_4$ is the total tetramer concentration, $\theta=(x_2-y_2)/z$ 
is the relative chirality of the dimers, $\phi=(x_4-y_4)/w$ is the 
relative chirality of tetramers.  Hence 
\beqa
\frac{\dd c_2}{\dd t} & = & - 2\mu c_2 + \mu\nu z - \alpha c_2 z , 
	\lbl{c23} \\ 
\frac{\dd z}{\dd t} & = & 2 \mu c_2 - \mu\nu z - \alpha c_2 z 
	- \xi z^2 (1+\theta^2) + 2 \beta w , \\ 
\frac{\dd w}{\dd t} & = & \alpha z c_2 + \half \xi z^2 (1+\theta^2) 
	- \beta w , \lbl{w3}\\ 
\frac{\dd \theta}{\dd t} & = & - \theta \left( \frac{2\mu c}{z} + 
	\frac{2\beta w}{z}+ \xi z (1-\theta^2) \right) + 
	\frac{2\beta w\phi}{z} , \\
\frac{\dd \phi}{\dd t} & = & \theta \frac{z}{w} ( \alpha c + \xi z ) 
	- \left( \alpha c + \half \xi z (1+\theta^2) \right) \frac{z}{w} \phi . 
\eeqa
The stability of the evolving symmetric-state ($\theta=\phi=0$) 
is given by the eigenvalues ($q$) of the matrix 
\beq
\left( \begin{array}{cc} 
- \left( \frac{2\mu c}{z} + \frac{2\beta w}{z} + \xi z \right) & 
\frac{2\beta w}{z} \\ 
(\alpha c + \xi z) \frac{z}{w} & 
- (\alpha c + \half \xi z) \frac{z}{w} 
\end{array} \right) , 
\eeq
which are given by
\beqa
q^2 + q \left( \frac{\alpha c z}{w} + \frac{\xi z^2}{w} 
+ \frac{2\mu c}{z} + \xi z + \frac{2\beta w}{z} \right) + && \nn \\ 
\frac{1}{w} \left( 2\mu c \alpha c + \mu c \xi z + 
\alpha c \xi z^2 + \half \xi^2 z^3 - \beta \xi z w \right) &=&0 . 
\eeqa
Hence there is an instability if 
\beq
\beta \xi z w > 2\mu c \alpha c + \mu c \xi z + 
\alpha c \xi z^2 + \half \xi^2 z^3 , 
\lbl{crude-instab}
\eeq
using the steady-state result that $2\beta w = z(2\alpha c + \xi z)$ 
and factorising ($2\alpha c + \xi z$) out of the result, reduces the 
instability (\ref{crude-instab}) to the contradictory $\xi z^2 > 
\xi z^2 + 2\mu c$. Hence the racemic steady-state of the system 
is stable for all choices of parameter values and is approached 
from all initial conditions. However, initial perturbations, 
may be amplified due to the presence of nonlinear terms. 

\begin{figure}[!ht]
\vspace*{75mm}
\includegraphics{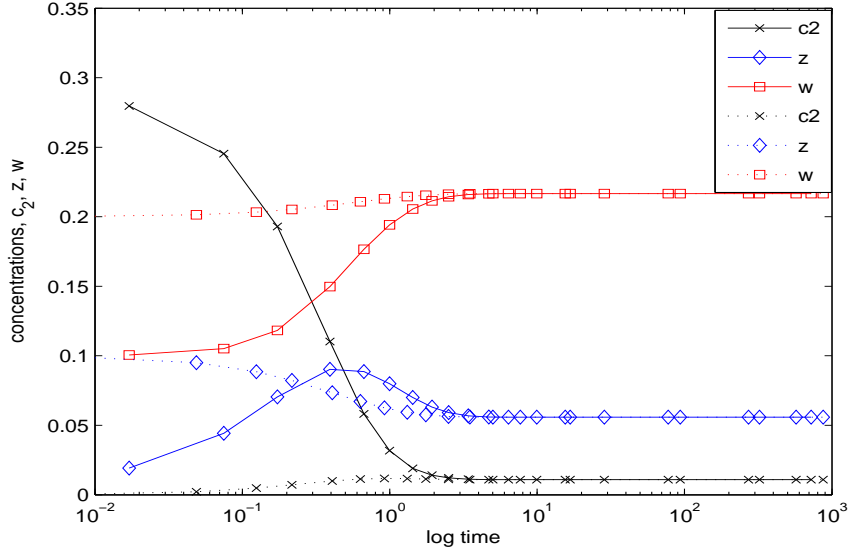}
\caption{The concentrations $c_2$, $z$ and $w$ 
(\protect\ref{tet-ztheta})--(\protect\ref{tet-wphi}) plotted against 
time, for the tetramer-truncated system with the two sets of initial 
data (\protect\ref{tet-ics}). Since model equations 
are in nondimensional form, the time units are arbitrary. The 
parameter values are $\mu=1$, $\nu=0.5$, $\alpha=\xi=10$, 
$\beta=0.1$. }
\label{fig-tet3}
\end{figure}

\begin{figure}[!ht]
\vspace*{75mm}
\includegraphics{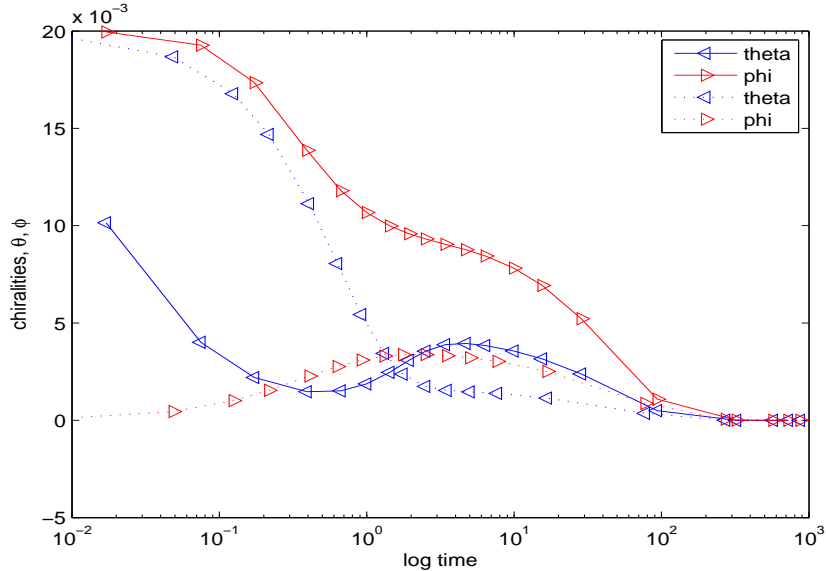}
\caption{The chiralities $\theta$, $\phi$ 
(\protect\ref{tet-ztheta})--(\protect\ref{tet-wphi}) plotted against 
time, for the tetramer-truncated system with 
the two sets of initial data (\protect\ref{tet-ics}). Since model equations 
are in nondimensional form, the time units are arbitrary. The 
parameter values are the same as in Figure \ref{fig-tet3}.}
\label{fig-tet4}
\end{figure}

Evolution from two sets of initial conditions of the system 
(\ref{dim-c2dot})--(\ref{dim-y4dot}) are shown in each of Figures 
\ref{fig-tet3}, \ref{fig-tet4}.   The continuous and dotted lines 
correspond to the initial data 
\beq \begin{array}{c} 
c_2(0) = 0.29 , \quad x_2(0) = 0.0051, \quad y_2(0) = 0.0049, \\ 
x_4(0) = 0.051 , \quad y_4(0) = 0.049 ; \quad {\rm and} \\ 
c_2(0) = 0 , \quad x_2(0) = 0.051 \quad y_2(0) = 0.049, \\  
x_4(0) = 0.1 , \quad y_4(0) = 0.1 ;  
\end{array} \lbl{tet-ics} \eeq
respectively. 
In the former case, the system starts with considerable amount 
of amorphous dimer, which is converted into clusters, and initially 
there is a slight chiral imbalance in favour of $x_2$ and $x_4$ over 
$y_2$ and $y_4$.  Over time this imbalance reduces (see figure 
\ref{fig-tet4}); although there is a region around $t=1$ where 
$\theta$ increases, both $\theta$ and $\phi$ eventually 
approach the zero steady-state.  

For both sets of initial conditions we note that the chiralities evolve 
over a significantly longer timescale than the concentrations, the 
latter having reached steady-state before $t=10$ and the former 
still evolving when $t={\cal O}(10^2)$.   In the second set of initial 
data, there is no $c_2$ present initially and there are exactly equal 
numbers of the two chiral forms of the larger cluster, but a slight 
exess of $x_2$ over $y_2$. In time an imbalance in larger clusters 
is produced, but over larger timescales, both $\theta$ and $\phi$ 
again approach the zero steady-state.  

Hence, we observe that the truncated system 
(\ref{dim-c2dot})--(\ref{dim-y4dot}) does {\em not} yield a chirally 
asymmetric steady-state.  Even though in the early stages of the 
reaction chiral perturbations may be amplified, at the end of the 
reaction there is a slower timescale over which the system returns 
to a racemic state. In the next section we consider a system 
truncated at hexamers to investigate whether that system allows 
symmetry-breaking of the steady-state. 

%-------------------------------------------------------------
\section{The truncation at hexamers}
\label{hex-sec}  \setcounter{equation}{0}

The above analysis has shown that the truncation of the model 
(\ref{gbd-c1})--(\ref{gbd-y2}) to (\ref{dim-c2dot})--(\ref{dim-y4dot}) 
results in a model which always ultimately approaches the 
symmetric (racemic) steady-state. In this section, we show that a 
more complex model, the truncation at hexamers retains enough 
complexity to demonstrate the symmetry-breaking bifurcation which 
occurs in the full system. In this case the governing equations are 
\beqa 
\ds\frac{\dd c_2}{\dd t} &=&  -2 \mu c_2 + \mu \nu (x_2+y_2) 
- \alpha c_2 (x_2+y_2) - \alpha c_2 (x_4+y_4) , 
\lbl{hex-c2} \\ 
\ds\frac{\dd x_2}{\dd t} & = & \mu c_2 - \mu \nu x_2 
- \alpha c_2 x_2 - 2 \xi x_2^2 - \xi x_2 x_4 + 2\beta x_4 
+ \beta x_6 , 
\\ 
\ds\frac{\dd x_4}{\dd t} & = & \alpha x_2 c_2 + \xi x_2^2 
- \beta x_4 - \alpha c_2 x_4 - \xi x_2 x_4 + \beta x_6 , 
\\
\ds\frac{\dd x_6}{\dd t} & = & \alpha x_4 c_2 + \xi x_2 x_4 
- \beta x_6 , 
\\
\ds\frac{\dd y_2}{\dd t} & = & \mu c_2 - \mu \nu y_2 
- \alpha c_2 y_2 - 2 \xi y_2^2 - \xi y_2 y_4 + 2\beta y_4 
+ \beta y_6 , 
\\ 
\ds\frac{\dd y_4}{\dd t} & = & \alpha y_2 c_2 + \xi y_2^2 
- \beta y_4 - \alpha c_2 y_4 - \xi y_2 y_4 + \beta y_6 , 
\\
\ds\frac{\dd y_6}{\dd t} & = & \alpha y_4 c_2 + \xi y_2 y_4 
- \beta y_6 . 
\lbl{hex-y6} \eeqa

To analyse the symmetry-breaking in the system we transform the 
dependent coordinates from $x_2,x_4,x_6,y_2,y_4,y_6$ to total 
concentrations $z,w,u$ and relative chiralities $\theta,\phi,\psi$ 
according to 
\beq \begin{array}{rclcrclcrcl}
x_2 &=& \half z (1 + \theta) , & \quad & 
x_4 &=& \half w (1 + \phi) , & \quad & 
x_6 &=& \half u (1 + \psi) , \\[2ex]   
y_2 &=& \half z (1 - \theta) , & \quad & 
y_4 &=& \half w (1 - \phi) , & \quad & 
y_6 &=& \half u (1 - \psi) . 
\end{array} \eeq

We now separate the governing equations for the total concentrations 
of dimers ($c,z$), tetramers ($w$) and hexamers ($u$)
\beqa
\ds\frac{\dd c}{\dd t} & = & - 2 \mu c 
+ \mu \nu z - \alpha c z - \alpha c w , 
\lbl{hex-cdot}\\ 
\ds\frac{\dd z}{\dd t} & =& 2\mu c - \mu \nu z 
- \alpha c z - \xi z^2 (1+\theta^2) - \half z w (1+\theta\phi) 
+ \beta u + 2 \beta w  , \nn \\ && 
\\ 
\ds\frac{\dd w}{\dd t} & = & \alpha c z  
+ \half \xi z^2 (1+\theta^2) - \beta w + \beta u 
- \alpha c w - \half \xi z w (1+\theta\phi)  , 
\\ 
\ds\frac{\dd u}{\dd t} & = & \alpha c w 
+ \half \xi z w (1+\theta\phi) - \beta u , 
\lbl{hex-udot}\eeqa
from those for the chiralities 
\beqa
\ds \frac{\dd \psi}{\dd t} & =& 
\frac{\alpha c w}{u} (\phi-\psi) + 
\frac{\xi z w}{2u} ( \theta+\phi-\psi-\psi\phi\theta )
\lbl{hex-psi-dot} \\ 
\ds \frac{\dd \phi}{\dd t} & = & 
\frac{\alpha c z }{w} (\theta-\phi) + 
\frac{\xi z^2}{2w} ( 2\theta -\phi-\phi\theta^2) + 
\frac{\beta u}{w} (\psi-\phi) - 
\half \xi z \theta (1-\phi^2) , \nn \\ && 
\\ 
\ds \frac{\dd \theta}{\dd t} & = & 
-\frac{2\mu c \theta}{z} - \xi z \theta(1\!-\!\theta^2) - 
\half \xi w \phi (1\!-\!\theta^2) + \frac{\beta u\psi}{z} - 
\frac{\beta u \theta}{z} \nn \\ && + \frac{2\beta w\phi}{z} 
- \frac{2\beta w \theta}{z} .   \lbl{hex-th-dot}
\eeqa

In applications, we expect $\nu<1$, so that the small amorphous 
clusters (dimers) prefer to adopt one of their chiral states rather 
than the achiral structure.   In addition, we note that the grinding 
process observed in experiments is much longer than the 
crystallisation process, and that there are many larger, macroscopic 
crystals hence we consider two limits in which $\beta \ll \alpha \xi$. 
We will consider the case of small $\beta$ with all other parameters 
being ${\cal O}(1)$ and then the case where $\alpha\sim\xi\gg1$ and 
all other parameters are ${\cal O}(1)$. 

%-------------------------
\subsection{Symmetric steady-state for the concentrations}

Firstly, let us solve for the symmetric steady-state.  
In this case we assume $\theta=0=\phi=\psi$, simplifying 
equations (\ref{hex-cdot})--(\ref{hex-udot}).   One of 
these is a redundant equation, hence we have the solution 
\beq
w = \frac{z}{\beta}(\alpha c + \half \xi z) , \qquad 
u = \frac{z}{\beta^2}(\alpha c+\half\xi z)^2 , 
\lbl{hex-wu-sol} \eeq
\beq
c = \frac{1}{\alpha} \left(\sqrt{ \left( \frac{\beta}{2} + 
\frac{\beta\mu}{\alpha z} + \frac{\xi z}{4} \right)^2 
+ \beta\mu\nu}  - \frac{\beta}{2} - \frac{\beta\mu}
{\alpha z} - \frac{\xi z}{4} \right) , 
\lbl{hex-c-sol} \eeq
with $z$ being determined by conservation of total mass in the system 
\beq 
2c + 2 z + 4 w + 6 u = \ro . \lbl{hex-roo} 
\eeq 

In the case of small grinding, ($\beta\ll1$), with $\ro$ 
and all other parameters being ${\cal O}(1)$, we find 
\beq  \begin{array}{rclcrcl}
z & = & \left( \ds\frac{2\ro \beta^2}{3 (\alpha\nu+\xi)^2} 
	\right)^{1/3} , &\qquad& 
c & = & \nu \left( \ds\frac{\ro \beta^2}{12 (\alpha\nu+\xi)^2} 
	\right)^{1/3} , \\ 
w & = & \left( \ds\frac{\ro^2 \beta}{18 (\alpha\nu+\xi)} 
	\right)^{1/3} , &\qquad& 
u & = & \ds\frac{\ro}{6} . 
\end{array}  \lbl{hex-ssss-asymp}  \eeq 
In this case most of the mass is in hexamers with a little in 
tetramers and very little in dimers. 

In the asymptotic limit of $\alpha \sim \xi \gg 1$ 
and all other parameters ${\cal O}(1)$, we find 
\beqa & 
c  = \ds\frac{\mu\nu}{\alpha} \left( 
\ds\frac{12\beta}{\ro\xi} \right)^{1/3} , \quad  
z = \left( \ds\frac{2\beta^2\ro}{3\xi^2} \right)^{1/3} , \quad 
w = \left( \ds\frac{\beta\ro^2}{18\xi} \right)^{1/3} ,  \quad
u = \ds\frac{\ro}{6} . 
& \nonumber \\ && \lbl{hex-sss2-asymp} 
\eeqa 
This differs significantly from the other asymptotic scaling as, not only 
are $c$ and $z$ both small, they are now different orders of magnitude, 
with $c\ll z$. We next analyse the stability of these symmetric states. 

%-------------------------
\subsection{Stability of symmetric state}

In deriving the above solutions (\ref{hex-wu-sol})--(\ref{hex-c-sol}), 
we have assumed chiral symmetry, that is, $\theta=0=\psi=\phi$.   
We now turn to analyse the validity of this assumption. Linearising 
the system of equations (\ref{hex-psi-dot})--(\ref{hex-th-dot}) 
which govern the chiralities, we determine whether the symmetric 
solution is stable from 
\beq \!\! 
\frac{\dd }{\dd t} \!\! \left( \!\! \begin{array}{c} 
\psi \\ \phi \\ \theta \end{array} \!\! \right) \!=\! 
\left( \begin{array}{ccc}  
\!\!\!- \ds\frac{\alpha c w}{u} \!-\! \ds\frac{\xi z w}{2u} \!\!& 
\ds\frac{\alpha c w}{u} \!+\! \ds\frac{\xi z w}{2u} & 
\ds\frac{\xi z w}{2u} \\[2ex] 
\ds\frac{\beta u}{w} & -\ds\frac{\alpha c z}{w} 
\!-\! \ds\frac{\xi z^2}{2w} \!-\! \ds\frac{\beta u}{w} & 
\ds\frac{\alpha c z}{w} \!+\! \ds\frac{\xi z^2}{w} 
\!-\! \half \xi z \\[2ex] 
\ds\frac{\beta u}{z} & 
\ds\frac{2\beta w}{z} \!-\! \ds\frac{\xi w}{2}  & \!\!
- \ds\frac{2\mu c}{z} \!-\! \xi z \!-\! \frac{\beta u}{z} 
\!-\! \ds\frac{2\beta w}{z} \!\!
\end{array} \! \right) \!\!
\left( \!\begin{array}{c} \psi \\ \phi \\ \theta 
\end{array} \!\right) \!.\!\!
\lbl{hex-stab-mat} \eeq
For later calculations it is useful to know the determinant of this matrix.  
Using the steady-state solutions (\ref{hex-wu-sol}), the determinant 
simplifies to 
\beq
D = \frac{3 c}{4 \beta \rho} ( 2 \alpha c + \xi z )^2 
( \alpha \xi z^2 - 4 \beta \mu ) . 
\eeq

For general parameter values, the signs of the real parts of the 
eigenvalues of the matrix in (\ref{hex-stab-mat}) are not clear. 
However, using the asymptotic result (\ref{hex-ssss-asymp}), 
for $\beta\ll1$, we obtain the simpler matrix 
\beq
\left( \!\!\begin{array}{ccc} 
-\beta & \beta & \ds \frac{\beta\xi}{\xi\!+\!\alpha\nu} 
\\[2ex] 
\left( \ds\frac{\beta^2 \ro (\xi\!+\!\alpha\nu) }{12} \right)^{1/3} & 
- \left( \ds\frac{\beta^2 \ro (\xi\!+\!\alpha\nu) }{12} \right)^{1/3} & 
-\frac{\xi}{2} \left( \ds\frac{2\beta^2\ro}{3(\xi\!+\!\alpha\nu)^2} \right) ^{1/3} 
\\[2ex]  
\beta^{1/3} \left( \ds\frac{\xi\!+\!\alpha\nu}{12\ro} \right)^{2/3} & 
- \frac{\xi}{2} \left( \ds\frac{\beta\ro^2}{18(\xi\!+\!\alpha\nu)} \right)^{1/3} & 
- \mu \nu - \beta^{1/3} \left( \ds\frac{\xi\!+\!\alpha\nu}{12\ro} \right)^{2/3} 
\end{array} \!\!\right) \! , \lbl{hex-asy1-mat}
\eeq
whose characteristic polynomial is 
\beq 
0 = q^3 + \mu\nu q^2 + \mu\nu \left( \rec{12} \beta^2 \ro 
(\xi\!+\!\alpha\nu) \right)^{1/3} q - D , \lbl{hex-asy1-cp} 
\eeq 
Formally $D$ is the determinant of the matrix in (\ref{hex-asy1-mat}), 
which is zero, giving a zero eigenvalue, which indicates marginal 
stability.   Hence, we return to the more accurate matrix in 
(\ref{hex-stab-mat}),  which gives $D \sim -\beta^2\mu\nu$.   
The polynomial (\ref{hex-asy1-cp}) thus has roots 
\beq 
q_1 \sim -\mu\nu, \quad 
q_2 \sim - \left( \frac{ \beta^2 \ro (\xi\!+\!\alpha\nu)}{12} \right)^{1/3} , 
\quad 
q_3 \sim - \left( \frac{12 \beta^4}{\ro(\alpha\nu\!+\!\xi)} \right)^{1/3} . 
\lbl{hex-ev1} \eeq 
This means that the symmetric state is always linearly stable 
for this asymptotic scaling.  We expect to observe evolution on 
three distinct timescales, one of ${\cal O}(1)$, one of 
${\cal O}(\beta^{-2/3})$ and one of ${\cal O}(\beta^{-4/3})$. 

We now consider the other asymptotic limit, namely, 
$\alpha\sim\xi\gg1$ and all other parameters are ${\cal O}(1)$.  
In this case, taking the leading order terms in each row, the stability 
matrix in (\ref{hex-stab-mat}) reduces to 
\beq
\left( \begin{array}{ccc} 
-6 \mu \nu \left( \frac{12\beta}{\ro\xi} \right)^{2/3} & 
6 \mu \nu \left( \frac{12\beta}{\ro\xi} \right)^{2/3} & 0
\\
\left( \frac{\beta^2\ro\xi}{12}\right)^{1/3} & 
-\left( \frac{\beta^2\ro\xi}{12}\right)^{1/3} & 
-\left( \frac{\beta^2\ro\xi}{12}\right)^{1/3} 
\\ 
\left( \frac{\beta\ro^2 \xi^2}{144} \right)^{1/3} &  
- \left( \frac{\beta\ro^2 \xi^2}{144} \right)^{1/3} &  
- \left( \frac{\beta\ro^2 \xi^2}{144} \right)^{1/3} 
\end{array} \right) , 
\eeq
which again formally has a zero determinant. 
The characteristic polynomial is 
\beq
0 = q^3 + q^2 + 6 \beta \mu\nu q - D , 
\eeq
wherein we again take the more accurate determinant 
obtained from a higher-order expansion of 
(\ref{hex-stab-mat}), namely $D=\beta^2\mu\nu$. 
The eigenvalues are then given by 
\beq
q_1 \sim - \left( \frac{\beta\ro^2\xi^2}{144} \right)^{1/3} , 
\qquad 
q_{2,3} \sim \pm \sqrt{\beta\mu\nu} 
\left( \frac{12\beta}{\ro\xi} \right)^{1/3} . 
\lbl{hex-ev2} \eeq
We now observe that there is always one stable and 
two unstable eigenvalues, so we deduce that the system 
breaks symmetry in the case $\alpha \sim \xi \gg 1$.  
The first eigenvalue corresponds to a faster timescale 
where $t\sim {\cal O}(\xi^{-2/3})$ whilst the latter two 
correspond to the slow timescale where $t={\cal O}(\xi^{1/3})$. 

%-------------------------
\subsection{Simulation results} 

\begin{figure}[!ht] 
\vspace*{64mm} 
\includegraphics{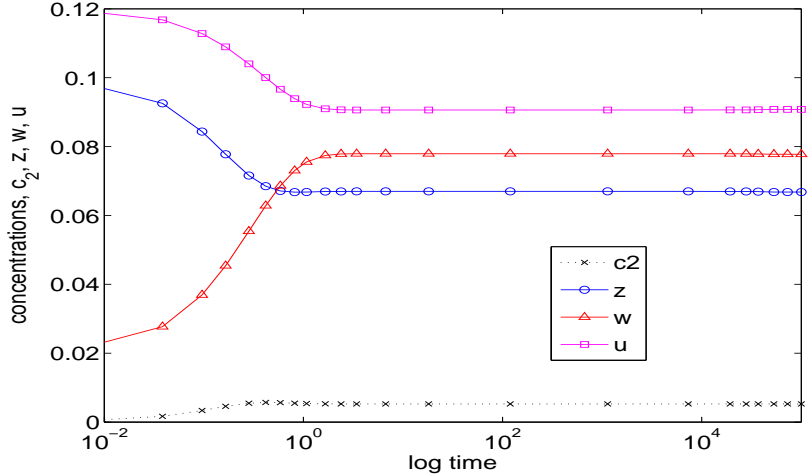} 
\caption{Illustration of the evolution of the total concentrations 
$c_2,z,w,u$ for a numerical solution of the system truncated 
at hexamers (\ref{hex-c2})--(\ref{hex-y6}) in the limit 
$\alpha\sim\xi \gg1$.  Since model equations 
are in nondimensional form, the time units are arbitrary. 
The parameters are $\alpha=\xi=30$, $\nu=0.5$, $\beta=\mu=1$, 
and the initial data is $x_6(0)=y_6(0)=0.06$, 
$x_4(0)=y_4(0)=0.01$, $x_2(0) = 0.051$, $y_2(0) = 0.049$, 
$c_2(0) = 0$. Note the time axis has a logarithmic scale. } 
\label{fig-hex-1} 
\end{figure}

\begin{figure}[!ht]
\vspace*{64mm}
\includegraphics{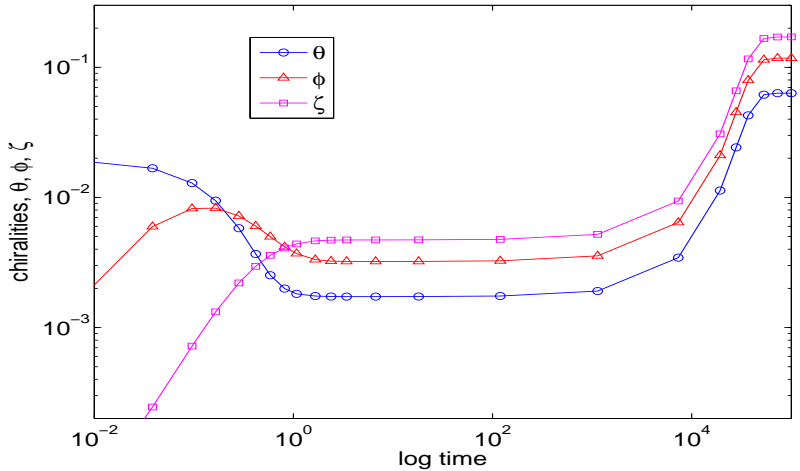}
\caption{Graph of the evolution of the chiralities against time 
on a log-log scale; results of numerical simulation of the same 
hexamer-truncated system, with identical initial data and 
parameters as in Figure \protect\ref{fig-hex-1}. }
\label{fig-hex-2}
\end{figure}

We briefly review the results of a numerical simulation of 
(\ref{hex-c2})--(\ref{hex-y6}) in the case $\alpha\sim\xi\gg1$ 
to illustrate the symmetry-breaking observed therein. 
Although the numerical simulation used the variables $x_k$ and 
$y_k$ ($k=2,4,6$) and $c_2$, we plot the total concentrations 
$z,w,u$ in Figure \ref{fig-hex-1}.   The initial conditions have a 
slight imbalance in the handedness of small crystals ($x_2,y_2$).  
The chiralities of small ($x_2,y_2,z$), medium ($x_4,y_4,w$), and 
larger ($x_6,y_6,u$) are plotted in Figure \ref{fig-hex-2} on a 
log-log scale. Whilst Figure \ref{fig-hex-1} shows the 
concentrations in the system has equilibrated by $t=10$, at this 
stage the chiralities are in a metastable state, that is, a long plateau 
in the chiralities between $t=10$ and $t=10^3$ where little appears 
to change.  There then follows a period of equilibration of chirality 
on the longer timescale when $t\sim 10^4$.   We have observed 
this significant delay between the equilibration of concentrations 
and that of chiralities in a large number of simulations.  The reason 
for this difference in timescales is due to the differences in the sizes 
of the eigenvalues in (\ref{hex-ev1}). 

We have also investigated the case $\beta \ll1$ with all other 
parameters ${\cal O}(1)$ to verify that this case does indeed approach 
the racemic state at large times (that is, $\theta,\phi,\zeta \rightarrow0$ 
as $t\rightarrow\infty$).   However, once again the difference in 
timescales can be observed, with the concentrations reaching 
equilibration on a faster timescale than the chiralities, due to the 
different magnitudes of eigenvalues (\ref{hex-ev2}).

%-------------------------------------------------------------
\section{New simplifications of the system}
\label{new-sec}
\setcounter{equation}{0}

We return to the equations (\ref{smm2})--(\ref{smmy2}) 
in the case $\delta=0$, now writing $x_2=x$ and $y=y_2$ 
to obtain 
\beqa
\frac{\dd c}{\dd t} & = & - 2 \mu c 
	+ \mu\nu (x+y) - \alpha c(N_x+N_y) , \lbl{newcdot} \\
\frac{\dd x}{\dd t} & = & \mu c - \mu\nu x - \alpha x c 
	+ \beta (N_x-x + x_4) - \xi x^2 - \xi x N_x , \lbl{newxdot} \\ 
\frac{\dd y}{\dd t} & = & \mu c - \mu\nu y  - \alpha y c 
	+ \beta (N_y-y + y_4) - \xi y^2 - \xi y N_y , \lbl{newydot}\\ 
\frac{\dd N_x}{\dd t} & = & \mu c - \mu\nu x 
	+ \beta (N_x-x) - \xi x N_x , \lbl{newnxdot} \\ 
\frac{\dd N_y}{\dd t} & = & \mu c - \mu\nu y 
	+ \beta (N_y-y) - \xi y N_y , \lbl{newnydot} 
\eeqa 
which are not closed, since $x_4,y_4$ appear 
on the {\sc rhs}'s of (\ref{newxdot}) and (\ref{newydot}), 
hence we need to find formulae to determine $x_4$ and $y_4$ 
in terms of $x,y,N_x,N_y$.

One way of achieving this is to expand the system to include other 
properties of the distribution of cluster sizes.  For example, equations 
governing the mass of crystals in each chirality can be derived as 
\beq
\frac{\dd \ro_x}{\dd t}=2\mu c-2\mu\nu x+2\alpha c N_x ,
\quad
\frac{\dd \ro_y}{\dd t}=2\mu c-2\mu\nu y+2\alpha c N_y . 
\lbl{new-roxy-dot} \eeq
These introduce no more new new quantities into the 
macroscopic system of equations, and do not rely on knowing 
$x_4$ or $y_4$, (although they do require knowledge of $x$ and $y$). 

In the remainder of this section we consider various potential formulae 
for $x_4$, $y_4$ in terms of macroscopic quantities so that a 
macroscopic system can be constructed.  We then analyse such  
macroscopic systems in two specific limits to show that predictions 
relating to symmetry-breaking can be made. 

%-------------------------------------------------
\subsection{Reductions}

The equations governing the larger cluster sizes $x_k$, $y_k$, are 
\beq 
\frac{\dd x_{2k}}{\dd t} = \beta( x_{2k+2} - x_{2k} ) 
- (x_{2k}-x_{2k-2})(\alpha c + \xi x) ;  
\lbl{5xkdot} \eeq 
in general this has solutions of the form $x_{2k} = \sum_j A_j(t) 
\Lambda_j^{k-1}$, 
where $\Lambda_j$ are parameters (typically taking values between 
unity (corresponding to a steady-state in which mass is being 
added to the distribution) and 
$\mfrac{\alpha c+\xi x}{\beta}$ (the equilibrium value); and $A_j(t)$ 
are time-dependent; for some $\Lambda_j$, $A_j$ will be constant. 

We assume that the distribution of each chirality of cluster is given by 
\beq 
x_{2k} = x \left( 1 - \frac{1}{\lambda_x} \right)^{k-1} ,\qquad\qquad
y_{2k} = y \left( 1 - \frac{1}{\lambda_y} \right)^{k-1} , 
\eeq
since solutions of this form may be steady-states of the governing 
equations (\ref{5xkdot}).  However, in our approximations for 
$x_4$ and $y_4$ the parameters $\lambda_x$, $\lambda_y$ are 
permitted to vary with time in some way that depends on other 
quantities in the model equations.   The resulting expressions for 
the macroscopic number and mass quantities are 
\beqa
N_x = \sum_{k=1}^\infty x_{2k} = x \lambda_x , &\qquad&  
N_y = \sum_{k=1}^\infty y_{2k} = y \lambda_y , \\   
\ro_x = \sum_{k=1}^\infty 2 k x_{2k} = 2 x \lambda_x^2 , &\qquad& 
\ro_y = \sum_{k=1}^\infty 2 k y_{2k} = 2 y \lambda_y^2 . 
\eeqa
Our aim is to find a simpler expression for the terms $x_4$ 
and $y_4$ which occur in (\ref{newxdot})--(\ref{newydot}), 
these are given by $x_4=x(1-1/\lambda_x)$ where 
\beq
\lambda_x = \frac{N_x}{x} = \frac{\ro_x}{2N_x} 
= \sqrt{\frac{\ro_x}{2x}} , \lbl{lambda-eqs}
\eeq
hence
\beq
x_4 = x - \frac{x^2}{N_x} , \quad 
x_4 = x - \frac{2 x N_x}{\ro_x} ,\quad 
{\rm or} \;\;\; 
x_4 = x - x\sqrt{\frac{2x}{\ro_x}} . 
\eeq

There are thus three possible reductions of the equations 
(\ref{newcdot})--(\ref{newnydot}), each eliminating one of 
$x,N_x,\ro_x$ (and the corresponding $y,N_y,\ro_y$). We consider 
each reduction in turn in the following subsections.  Since some 
of these reductions involve $\ro_x, \ro_y$, we also use the 
evolution equations (\ref{new-roxy-dot}) for these quantities. 

%-------------------------------------------------
\subsection{Reduction 1: to $x,y,N_x,N_y$}

Here we assume $\lambda_x = N_x/x$, $\lambda_y = N_y/y$, 
so, in addition to (\ref{newcdot}), (\ref{newnxdot})--(\ref{newnydot}) 
the equations of motion are 
\beqa
\frac{\dd x}{\dd t} & = & \mu c - \mu \nu x + \beta N_x 
	- \frac{\beta x^2}{N_x} - \xi x^2 - \xi x N_x ,  \\
\frac{\dd y}{ \dd t} & = & \mu c - \mu \nu y + \beta N_y 
	- \frac{\beta y^2}{N_y} - \xi y^2 - \xi y N_y ; 
\eeqa
we have no need of the densities $\ro_x,\ro_y$ in this formulation. 

The disadvantage of this reduction is that, due to 
(\ref{lambda-eqs}), the total mass is given by 
\beq
\ro = 2c + \ro_x+\ro_y = 2 c + \frac{2 N_x^2}{x} 
+ \frac{2 N_y^2}{y} , 
\eeq
and there is no guarantee that this will be conserved. 

We once again consider the system in terms of total concentrations 
and relative chiralities by applying the transformation 
\beqa&
x = \half z (1\!+\!\theta) , \quad 
y = \half z (1\!-\!\theta) , \quad 
N_x = \half N (1\!+\!\phi) , \quad 
N_y = \half N (1\!-\!\phi) , & \nn \\ && 
\eeqa
to obtain the equations
\beqa
\frac{\dd c}{\dd t} & =& - 2 \mu c + \mu \nu z - \alpha c N , 
\lbl{R1cdot} \\ 
\frac{\dd z}{\dd t} & =& 2\mu c - \mu \nu z - \alpha c z + \beta N 
	-\frac{\beta z^2(1+\theta^2-2\theta\phi)}{N(1-\phi^2)} \nn \\ && 
	- \half \xi z^2(1+\theta^2) - \half \xi z N (1+\theta\phi) , 
	\lbl{R1zdot} \\
\frac{\dd N}{\dd t} & = & 2\mu c - \mu \nu z + \beta N - \beta z 
	- \half \xi z N (1+\theta\phi) . \lbl{R1Ndot} \\ 
\frac{\dd \theta}{\dd t} & = & 
	- \left( \mu \nu + \alpha c + \xi z + \half \xi N 
		+ \frac{2\beta z}{N(1\!-\!\phi^2)} 
		+ \frac{1}{z} \frac{\dd z}{\dd t} \right) \theta \nn \\ && 
	+ \left( \frac{\beta N}{z} - \half \xi N 
	+ \frac{\beta z (1\!+\!\theta^2)}{N(1\!-\!\phi^2)} 
	\right) \phi , \\ 
\frac{\dd \phi}{\dd t} & =& 
	- \left( \mu\nu + \beta + \half \xi N \right) \frac{z}{N} \theta 
	+ \left( \beta - \half \xi z - \frac{1}{N}\frac{\dd N}{\dd t} \right) \phi . 
	\nn \\ && 
\eeqa
These equations have the symmetric steady-state given by 
$\theta=0=\phi$ and $c,z,N$ satisfying 
\beq
c = \frac{\mu\nu z}{2\mu+\alpha N}  , \qquad
z = \frac{2\beta N (2\mu+\alpha N) }
{(2\beta + \xi N)(2\mu+\alpha N)+2\alpha\mu\nu N} , 
\lbl{R1sss} \eeq
from (\ref{R1cdot}) and (\ref{R1Ndot}).   Note that the steady 
state value of $N$ will depend upon the initial conditions, it is 
not determined by (\ref{R1zdot}). This is because the 
steady-state equations obtained by setting the time derivatives 
in (\ref{R1cdot})--(\ref{R1Ndot}) are not independent.  
The difference (\ref{R1zdot})--(\ref{R1Ndot}) is equal 
to $z/N$ times the sum (\ref{R1cdot})$+$(\ref{R1Ndot}).  

In subsections \ref{5R1A1-sec} and \ref{5R1A2-sec} below, 
so as to discuss the stability of a solution in the two asymptotic 
regimes $\beta\ll1$ and $\alpha\sim\xi\gg1$, we augment the 
steady-state equations (\ref{R1cdot})--(\ref{R1Ndot}) with the 
condition $\ro=2N^2/z$, with $\ro$ assumed to be ${\cal O}(1)$.

The linear stability of $\theta=0=\phi$ is given by assuming 
$\theta$ and $\phi$ are small, yielding the system 
\beq 
\frac{\dd}{\dd t} \!\!\left(\!\! \begin{array}{c} 
\theta \\[2ex] \phi \end{array}\!\! \right) \!
= \! \left( \begin{array}{cc} 
- \left( \ds\frac{2\mu c}{z} + \ds\frac{\xi z}{2} 
	+ \ds\frac{\beta z}{N} 
	+ \ds\frac{\beta N}{z} \right) &
\left(\ds\frac{\beta N}{z} + \ds\frac{\beta z}{N} 
	- \ds\frac{\xi N}{2} \right) \\ 
- ( \mu \nu + \beta + \half \xi N ) \ds\frac{z}{N} & 
\left( \beta + \mu\nu - \ds\frac{2\mu c}{z} \right) 
\ds\frac{z}{N} \end{array} \right) \!\! \left( \!\!
\begin{array}{c} \theta \\[2ex] \phi \end{array} \!\!\right)\! . 
\lbl{5stabmat} \eeq
An instability of the symmetric solution is indicated by the 
determinant of this matrix being negative. Substituting (\ref{R1sss}) 
into the determinant, yields 
\beq
\mbox{det} = \frac{ \beta \mu \nu ( 4 \beta \mu - \alpha \xi N^2 )}
{4\beta\mu + 2 \alpha \beta N + 2 \mu \xi N + 2 \alpha \mu \nu N 
+ \alpha \xi N^2} . 
\lbl{5detsimp} \eeq
Hence we find that the symmetric (racemic) state is unstable if 
$N > 2 \sqrt{ \mu\beta / \alpha \xi }$, that is,  large aggregation rates 
($\alpha,\xi$) and slow grinding ($\beta$) are preferable for 
symmetry-breaking. 

We consider two specific asymptotic limits of parameter values 
so as to derive specific results for steady-states and conditions on 
stability.  In both limits, we have that the aggregation rates dominate 
fragmentation ($\alpha \sim \xi \gg \beta$), so that the system is 
strongly biased towards the formation of crystals and the dimer 
concentrations are small.   In the first case we assume that the 
fragmentation is small and the aggregation rates are of a similar 
scale to the interconversion of dimers ($\beta \ll \mu \sim \alpha \sim 
\xi = {\cal O}(1)$); whilst the second has a fragmentation rate of 
similar size to the dimer conversion rates and larger aggregation rates 
($\alpha \sim \xi \gg \mu \sim \beta = {\cal O}(1)$). 

%------------------
\subsubsection{Asymptotic limit 1: $\beta\ll1$} 
\label{5R1A1-sec} 

In the case of  asymptotic limit 1, $\beta\ll1$, we find the 
steady-state solution 
\beq
N \sim \sqrt{\frac{\beta\ro}{\xi+\alpha\nu}} , \quad 
z \sim \frac{2\beta}{\xi+\alpha\nu} , \quad 
c \sim \frac{\beta\nu}{\xi+\alpha\nu} . 
\eeq
From (\ref{5detsimp}), we find an instability if $\ro > \ro_c := 
4 \mu (\xi+\alpha\nu) / \alpha\xi$. That is, larger masses ($\ro$) 
favour symmetry-breaking, as do larger aggregation rates 
($\alpha,\xi$).  The eigenvalues of (\ref{5stabmat}) in this limit are 
$q_1 = -\mu\nu$ -- a fast stable mode of the dynamics and 
\beq
q_2 = \frac{\alpha \xi \beta^{3/2}}{2\mu \sqrt{\ro} (\xi+\alpha\nu)^{3/2}} 
\left( \ro - \frac{4\mu(\xi+\alpha\nu)}{\alpha\xi} \right) , 
\eeq
which indicates a slowly growing instability when $\ro>\ro_c$.  Hence 
the balace of achiral to chiral morphologies of smaller clusters ($\nu$) 
also influences the propensity for non-racemic solution.  However, 
since the dynamics described by this model does not conserve total 
mass, the results from this should be treated with some caution, 
and we now analyse models which do conserve total mass. 

%------------------
\subsubsection{Asymptotic limit 2: $\alpha\sim\xi\gg1$}
\label{5R1A2-sec}

In this case we find the steady-state solution is given by 
\beq
N \sim \sqrt{\frac{\beta\ro}{\xi}} , \quad  
z \sim \frac{2\beta}{\xi} , \quad 
c \sim \frac{4\mu\nu}{\alpha} \sqrt{\frac{\beta}{\xi\ro}} . 
\eeq
The condition following from (\ref{5detsimp}) then implies that we 
have an instability if $\ro>\ro_c := 4\mu/\alpha \ll 1$. The eigenvalues 
of the stability matrix are $q_1 = - \half \sqrt{\beta\ro\xi}$, which is 
large and negative, indicating attraction to some lower dimensional 
solution over a relatively fast timescale; the eigenvector being 
$(1,0)^T$ showing that $\theta\rightarrow0$. The other eigenvalue 
is $q_2 = 2\mu\nu \sqrt{\beta/\ro\xi} \ll 1$, and  corresponds to a slow 
growth of the chirality of the solution, since it relates to the 
eigenvector $(0,1)^T$. Assuming the system is initiated near its 
symmetric solution ($\theta=\phi=0$), this shows that the distribution 
of clusters changes its chirality first, whilst the dimer concentrations 
remain, at least to leading order, racemic.  We expect that at a later 
stage the chirality of the dimers too will become nonzero. 

%-------------------------------------------------
\subsection{Reduction 2: to $x,y,\ro_x,\ro_y$}

Here we eliminate $x_4=x(1-1/\lambda_x)$, 
$y_4=y(1-1/\lambda_y)$ together with $N_x$ and $N_y$ using 
\beq
\lambda_x=\sqrt{\frac{\ro_x}{2x}}, \quad 
\lambda_y=\sqrt{\frac{\ro_y}{2y}}, \quad 
N_x = \sqrt{\frac{x\ro_x}{2}}, \quad 
N_y = \sqrt{\frac{y\ro_y}{2}}, 
\eeq
leaving a system of equations for $(c,x,y,\ro_x,\ro_y)$ 
\beqa
\frac{\dd c}{\dd t} & = & \mu\nu(x+y) - 2\mu c 
- \sqrt{2} \alpha c \left( \sqrt{x\ro_x} + \sqrt{y \ro_y} \right) , \\ 
\frac{\dd x}{\dd t} & =& \mu c - \mu \nu x - \alpha c x - 
\xi x^2 - \xi x \sqrt{\frac{x\ro_x}{2}} + \beta \sqrt{\frac{x\ro_x}{2}} 
- \beta x \sqrt{\frac{2x}{\ro_x}} , \nn \\ && \\ 
\frac{\dd \ro_x}{\dd t} & = & - 2 \mu \nu x + 2 \mu c 
+ 2 \alpha c \sqrt{\frac{x\ro_x}{2}} , 
\eeqa 
with similar equations for $y,\ro_y$.  Transforming to total 
concentrations and relative chiralities by way of 
\beqa&
x = \half z (1+\theta) , \quad 
y = \half z (1-\theta) , \quad 
\ro_x = \half R (1+\zeta) , \quad 
\ro_y = \half R (1-\zeta) , 
&\nn\\&&
\eeqa
we find 
\beqa
\frac{\dd c}{\dd t} & =& \mu \nu z - 2 \mu c 
	- \frac{\alpha c \sqrt{z R}}{2\sqrt{2}} \left[ 
	\sqrt{(1\!+\!\theta)(1\!+\!\zeta)} + 
	\sqrt{(1\!-\!\theta)(1\!-\!\zeta)} \right] , \lbl{new-r2-cdot} 
\\ 
\frac{\dd z}{\dd t} & = & 2\mu c - \mu \nu z - \alpha c z 
	- \half \xi z^2 (1\!+\!\theta^2) \nn \\ && 
	+ \frac{\beta \sqrt{zR}}{2\sqrt{2}} 
	\left[ \sqrt{(1\!+\!\theta)(1\!+\!\zeta)} + \sqrt{(1\!-\!\theta)(1\!-\!\zeta)} 
	\right] \nn \\ && 
	- \frac{\xi z^{3/2} R^{1/2}}{4\sqrt{2}} \left[ 
	(1\!+\!\theta)^{3/2} (1\!+\!\zeta)^{1/2} + (1\!-\!\theta)^{3/2} 
	(1\!-\!\zeta)^{1/2} \right] \nn \\ && 
	- \frac{\beta z^{3/2} }{\sqrt{2R}} 
	\left[ \frac{(1\!+\!\theta)^{3/2}}{(1\!+\!\zeta)^{1/2}} + 
	\frac{(1\!-\!\theta)^{3/2}}{(1\!-\!\zeta)^{1/2}} \right] , 
\lbl{new-r2-zdot} 
\\ 
\frac{\dd R}{\dd t} & = & - 2\mu\nu z + 4 \mu c 
	+ \half \alpha c \sqrt{2zR} \left[ 
	\sqrt{(1\!+\!\theta)(1\!+\!\zeta)} + 
	\sqrt{(1\!-\!\theta)(1\!-\!\zeta)} \right] , 
\nn \\ && \lbl{new-r2-Rdot} 
\eeqa
together with the equations 
(\ref{new-r2-thetadot})--(\ref{new-r2-zetadot}) for the relative 
chiralities $\theta$ and $\zeta$,  which will be analysed later. 

Since the equations for $\dd R/dd t$ and $\dd c/\dd t$ are 
essentially the same, we obtain a third piece of information from 
the requirement that the total mass in the system is unchanged 
from the initial data, hence the new middle equation above. 
Solving these we find $c=\half (\ro-R)$ and use this in place 
of the equation for $c$. 

In the symmetric case ($\theta=\zeta=0$) we obtain the 
steady-state conditions 
\beqa
0 & = & 2\mu\nu z - 4\mu c - \alpha c \sqrt{2zR} , \qquad\qquad 
\ro \; = \; R + 2 c , \lbl{R2-ssss1} \\ 
0 & = & 2\mu c - \mu \nu z - \alpha c z - \half \xi z^2 
	+ \half \beta \sqrt{2zR} - \beta z \sqrt{\frac{2z}{R}} 
	- \frac{\xi z}{2} \sqrt{\frac{zR}{2}} . \nn \\ && \lbl{R2-ssss2} 
\eeqa
For small $\theta,\zeta$, the equations for the chiralities 
can be approximated by 
\beqa
\frac{\dd \theta}{\dd t} & = & - \left( \frac{2\mu c}{z} 
	+ \half \xi z + \half \beta \sqrt{\frac{R}{2z}} 
	+ \half \beta \sqrt{\frac{2z}{R}} 
	+ \rec{4} \xi \sqrt{\frac{zR}{2}} \right) \theta \nn \\ && 
	+ \left( \frac{\beta(R+2z)}{2\sqrt{2zR}} 
	- \frac{\xi}{4} \sqrt{\frac{Rz}{2}} \right) \zeta , 
\lbl{new-r2-thetadot}  \\ 
\frac{\dd \zeta}{\dd t} & = & \left( \frac{2\mu\nu z}{R} 
- \alpha c \sqrt{\frac{zR}{2}} \right) \theta 
- \left( \frac{2\mu\nu z}{R} - \frac{4\mu c}{R} \right) \zeta , 
\lbl{new-r2-zetadot} 
\eeqa
We analyse the stability of the symmetric (racemic) state in the two 
limits $\beta\ll1$ and $\alpha\sim\xi\gg1$ in the next subsections. 

%-------------------------------------------------
\subsubsection{Asymptotic limit 1: $\beta\ll1$}

In this case, solving the conditions 
(\ref{R2-ssss1})--(\ref{R2-ssss2}) asymptotically, 
we find 
\beq
z \sim \frac{2\beta}{\xi+\alpha\nu} , \qquad 
c \sim \frac{\beta\nu}{\xi+\alpha\nu} , \qquad 
R \sim \ro - 2c . 
\eeq
Substituting these values into the differential equations 
which determine the stability of the racemic state leads to 
\beq
\frac{\dd }{\dd t} 
\left( \begin{array}{c} \theta \\[3ex] \zeta \end{array} \right) 
\left( \begin{array}{cc} 
-\mu\nu & 
\ds\frac{\alpha\nu}{4} \sqrt{\ds\frac{\beta\ro}{\xi+\alpha\nu}}\\ 
-\ds\frac{4\beta\mu\nu}{\ro(\xi+\alpha\nu)} & 
\ds\frac{\alpha\nu\beta^{3/2}}{(\xi+\alpha\nu)^{3/2} \sqrt{\ro}}  
\end{array} \right) 
\left( \begin{array}{c} \theta \\[3ex] \zeta \end{array} \right) . 
\eeq
Formally this matrix has eigenvalues of zero and $-\mu\nu$. 
Since the zero eigenvalue indicates marginal stability of the 
racemic solution, we need to consider higher-order terms to 
obtain a more definite result. 

Going to higher order, gives the determinant of the resulting matrix 
as $-\alpha \xi \nu / (\alpha\nu+\xi)^2$ hence the eigenvalues are 
\beq
q_1 = -\mu\nu , \qquad {\rm and} \quad 
q_2 = \frac{ \alpha \xi }{\mu (\alpha\nu+\xi)^2 } , 
\eeq
the former indicating a rapid decay of $\theta$ (corresponding to the 
eigenvector $(1,0)^T$), and the latter showing a slow divergence from 
the racemic state in the $\zeta$-direction, at leading order, according to 
\beq
\left( \begin{array}{c} \theta \\ \zeta \end{array} \right) 
\sim C_1 \left( \begin{array}{c} 0 \\ 1 \end{array} \right) 
\exp \left( \frac{ \alpha \xi t }{\mu (\alpha\nu+\xi)^2 } \right) . 
\eeq
Hence in the case $\beta\ll1$, we find an instability of the 
symmetric solution for all other parameter values. 

%-------------------------------------------------
\subsubsection{Asymptotic limit 2: $\alpha\sim\xi\gg1$}

In this case, solving the conditions 
(\ref{R2-ssss1})--(\ref{R2-ssss2}) asymptotically, we find 
\beq 
z \sim \frac{2\beta}{\xi} , \qquad 
c \sim \frac{2\mu\nu}{\alpha} \sqrt{\frac{\beta}{\ro\xi}} , \qquad 
R \sim \ro - 2c . 
\eeq 
Substituting these values into the differential equations 
(\ref{new-r2-thetadot})--(\ref{new-r2-zetadot}) 
which determine the stability of the racemic state leads to 
\beq 
\frac{\dd }{\dd t} 
\left( \begin{array}{c} \theta \\[1ex] \zeta \end{array} \right) 
\left( \begin{array}{ccc} 
- \half \sqrt{\beta\xi\ro} && o(\sqrt{\xi}) \\[1ex] 
- \ds\frac{4\beta\mu\nu}{\ro\xi} && \ds\frac{4\beta\mu\nu}{\ro\xi} 
\end{array} \right) 
\left( \begin{array}{c} \theta \\[1ex] \zeta \end{array} \right) , 
\eeq 
hence the eigenvalues are $q_1=-\half\sqrt{\beta\ro\xi}$ and 
$q_2 = 4\mu\nu\beta/\ro\xi$, (in the above $o(\sqrt{\xi})$ means 
a quantity $q$ satisfying $q\ll\sqrt{\xi}$ as $\xi\rightarrow\infty$).  
Whilst the former indicates the existence of a stable manifold (with 
a fast rate of attraction), the latter shows that there is also an unstable 
manifold. Although the timescale associated with this is much slower, 
it shows that the symmetric (racemic) state is unstable. 

%-------------------------------------------------
\subsection{Reduction 3: to $N_x,N_y,\ro_x,\ro_y$}

In this case our aim is to retain only information on the number 
and typical size of crystal distribution, so we eliminate the dimer 
concentrations $x,y$, using 
\beq 
\lambda_x = \frac{\ro_x}{2 N_x} , \quad 
\lambda_y = \frac{\ro_y}{2 N_y} , \quad 
x = \frac{2 N_x^2}{\ro_x} , \quad 
y = \frac{2 N_y^2}{\ro_y} . 
\eeq 
These transformations reformulate the governing equations 
(\ref{newcdot})--(\ref{new-roxy-dot}) to 
\beqa
\frac{\dd N_x}{\dd t} & = & \half \mu (\ro -R) + \beta N_x  
	- 2 (\mu\nu+\beta) \frac{N_x^2}{\ro_x} 
	- \frac{2\xi N_x^3}{\ro_x} , \lbl{r3-nxdot} \\ 
\frac{\dd N_y}{\dd t} & = & \half \mu (\ro - R) + \beta N_y
	- 2 (\mu\nu+\beta) \frac{N_y^2}{\ro_y} 
	- \frac{2\xi N_y^3}{\ro_y} , \\ 
\frac{\dd \ro_x}{\dd t} & = & (\ro-R)(\mu+\alpha N_x) 
	- \frac{4\mu\nu N_x^2}{\ro_x} , \lbl{r3-roxdot} \\ 
\frac{\dd \ro_y}{\dd t} & = & (\ro-R)(\mu+\alpha N_y) 
	- \frac{4\mu\nu N_y^2}{\ro_y} , \lbl{r3-roydot}
\eeqa
where $R := \ro_x + \ro_y$. 
We now transform to total concentrations ($N$, $R$) 
and relative chiralities ($\phi$ and $\zeta$) {\em via}  
\beq 
N_x = \half N (1+\phi) , \quad 
N_y = \half N (1-\phi) , \quad 
\ro_x = \half R (1+\zeta) , \quad 
\ro_y = \half R (1-\zeta) , 
\eeq
together with $c = \half (\ro - R)$, to obtain 
\beqa
\frac{\dd R}{\dd t} & = & (\ro-R)(2\mu+ \alpha N) 
- \frac{4\mu\nu N^2(1+\phi^2-2\phi\zeta)}{R (1-\zeta^2)} , 
\lbl{r3Rd} \\ \lbl{r3Nd}  
\frac{\dd N}{\dd t} & = & \!\!\mu (\ro \! - \! R) + \beta N 
	\\ && \! - \frac{N^2}{R(1\!-\!\zeta^2)} \left[ 
	2(\mu\nu\!+\!\beta) (1\!+\!\phi^2\!-\!2\phi\zeta) + 
	\xi N (1\!+\!3\phi^2\!-\!3\phi\zeta\!-\!\phi^3\zeta) \right] ,
\nn \\ 
\frac{\dd\phi}{\dd t} &=& \beta\phi - \frac{1}{N}\frac{\dd N}{\dd t}\phi
	\\&& \!\!- \frac{N}{R(1\!-\!\zeta^2)} \left[ 
	2(\beta\!+\!\mu\nu)(2\phi\!-\!\zeta\!-\!\phi^2\zeta)
	+ \xi N (3\phi\!-\!\zeta\!+\!\phi^3\!-\!3\phi^2\zeta) \right] , \nn 
\\ 
\frac{\dd \zeta}{\dd t} & =& \frac{\alpha (\ro-R) N \phi}{R} 
	- \frac{1}{R}\frac{\dd R}{\dd t} \zeta - \frac{4\mu\nu N^2 
	(2\phi-\zeta-\phi^2\zeta)}{R^2 (1-\zeta^2)} . 
\eeqa
We now analyse this system in more detail, since this set of 
equations conserves mass, and is easier to analyse than 
(\ref{new-r2-cdot})--(\ref{new-r2-Rdot}) due to the absence of 
square roots.  We consider the two asymptotic limits ($\beta\ll1$ 
and $\alpha\sim\xi\gg1$) in which, at steady-state, the majority of 
mass is in the form of clusters. 

%----------------------
\subsubsection{The symmetric steady-state}

Putting $\zeta=0=\phi$, we find the symmetric steady-state is given by 
\beqa 
0 &=& (\ro-R)(2\mu+\alpha N) - \frac{4\mu\nu N^2}{R} , 
\lbl{r3ssss1} \\ 
0 &=& \mu (\ro-R) + \beta N 
- 2(\mu\nu+\beta)\frac{N^2}{R} - \frac{\xi N^3}{R} . 
\lbl{r3ssss2} \eeqa 
the former is solved by one of 
\beq
R = \half \ro \left( 1 \pm \sqrt{ 1 - \frac{16\mu\nu N^2} 
{ (2\mu+\alpha N) \ro^2 } } \right) , \qquad 
\eeq
\beq
N = \frac{\alpha R(\ro-R)}{8\mu\nu} \left( 1 + 
\sqrt{1 + \frac{32\mu^2\nu}{\alpha^2 R(\ro-R)}} \right) . 
\eeq
More complete asymptotic solutions will be derived in Sections 
\ref{r3-a1-sec} and \ref{r3-a2-sec}.  

%----------------------
\subsubsection{Stability of the symmetric state} 

We now consider the stability of the 
symmetric steady-state. For small $\phi,\zeta$ we have 
\beqa \!\!\!\!\!& 
\ds\frac{R}{N} \ds\frac{\dd}{\dd t} \!\!
\left( \!\!\begin{array}{c} \phi \\ \\ \zeta \end{array} \!\!\right) 
\!=\!\! \left( \!\!\begin{array}{cc} 
\!\! - \! 2\beta \!-\! 2\mu\nu \!-\! 2 \xi N 
	\!-\! \ds\frac{\mu (\ro\!-\!R) R}{N^2} \!\!&\!  
2\beta \!+\! 2\mu\nu \!+\! \xi N   
\\  
\left( \alpha (\ro\!-\!R) 
	\!-\! \ds\frac{8\mu\nu N}{R} \right) \! \!&\! 
\!8\mu\nu \!-\! \ds\frac{(\ro\!\!-\!\!R)(2\mu\!\!+\!\!\alpha N)R}{N^2} \!  
\end{array} \!\!\right) \!\!\!
\left(\!\! \begin{array}{c} \phi \\ \\ \zeta \end{array} \!\!\right) \!\!,\!\! 
& \nn \\ \!\!\!\!\!\! && \!\!\!\!\lbl{r3-stab}
\eeqa
and this is unstable if the determinant of this matrix is negative. 
Now we consider the two asymptotic limits in more detail. 

%----------------------
\subsubsection{Asymptotic limit 1: $\beta \ll1$}
\label{r3-a1-sec}

When fragmentation is slow, that is, $\beta\ll1$, at steady-state we 
have $N={\cal O}(\sqrt{\beta})$ and $R = \ro - {\cal O}(\beta)$. 
Balancing terms in (\ref{r3ssss1})--(\ref{r3ssss2}) we find the same 
leading order equation twice, namely $2\nu N^2=\beta\ro(\ro-R) $. 
Taking the difference of the two yields an independent equation 
from higher order terms, hence we obtain 
\beq 
N \sim \sqrt{\frac{\beta \ro}{\xi+\alpha\nu}} , \qquad 
R \sim \ro - \frac{2\nu\beta}{\xi+\alpha\nu} . 
\eeq 
Note that this result implies that the dimer concentrations 
are small, with $c\sim z$ and $c \sim \beta\nu / (\xi+\alpha\nu)$, 
$z\sim 2\beta/(\xi+\alpha\nu)$.  

\begin{figure}[!ht]
\vspace*{68mm}
\includegraphics{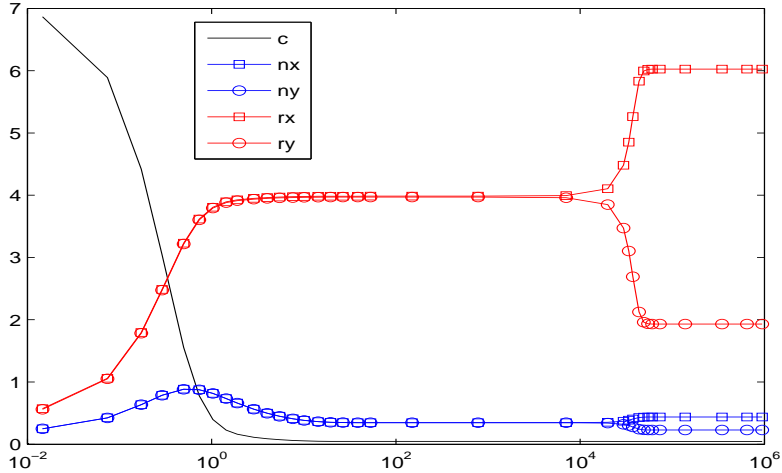}
\caption{Graph of concentrations $N_x,N_y,\ro_x,\ro_y,c$ 
against time on a logarithmic time for the asymptotic limit 1, 
with initial conditions $N_x=0.2=N_y$, $\ro_x=0.45$, $\ro_y=0.44$, 
other parameters given by $\alpha=1=\xi=\mu$, $\beta=0.01$ , 
$\ro=8$. Since model equations are in nondimensional form, 
the time units are arbitrary.  }
\label{fig-r3alpha}
\end{figure}

Substituting these expressions into those for the stability of the 
symmetric steady-state (\ref{r3-stab}), we find 
\beq
\frac{R}{4\mu\nu N} \frac{\dd}{\dd t} 
\left( \begin{array}{c} \phi \\[1ex] \zeta \end{array} \right) = 
\left( \begin{array}{cc} -1 & \quad \frac{1}{2} \\ 
-2\sqrt{\ds\frac{\beta}{\ro(\xi\!+\!\alpha\nu)}} & \quad 1 
\end{array} \right) 
\left( \begin{array}{c} \phi \\[1ex] \zeta \end{array} \right) . 
\eeq
This matrix has one stable eigenvalue (corresponding to 
$(1,0)^T$ and hence the decay of $\phi$ whilst $\zeta$ remains 
invariant), the unstable eigenvector is $(1,4)^T$, hence we find 
\beq
\left( \begin{array}{c} \phi(t) \\ \zeta(t) \end{array} \right) \sim C 
\left( \begin{array}{c} 1 \\ 4 \end{array} \right) \exp \left( 
\frac{4\mu\nu t \sqrt{\beta}}{\sqrt{\ro(\xi+\alpha\nu)}} \right) . 
\lbl{r3-chir-rate} \eeq 
If we compare the timescale of this solution to that over which the 
concentrations $N,R$ vary, we find that symmetry-breaking 
occurs on a slower timescale than the evolution of cluster masses 
and numbers. This is illustrated in the numerical simulation of 
equations (\ref{r3-nxdot})--(\ref{r3-roydot}) shown in Figure 
\ref{fig-r3alpha}.  More specifically, the time-scale increases with 
the mass in the system, and with the ratio of aggregation to 
fragmentation rates, $(\alpha\nu+\xi)/\beta$, and is inversely related 
to the chiral switching rate of small clusters ($\mu\nu$). 

%----------------------
\subsubsection{Asymptotic limit 2: $\alpha \sim \xi \gg 1$}
\label{r3-a2-sec}

In this case we retain the assumptions that $\mu,\nu={\cal O}(1)$, 
however, we now impose $\beta={\cal O}(1)$ and 
$\alpha \sim \xi \gg1$.  For a steady-state, we require the scalings 
$N ={\cal O}(1/\sqrt{\xi})$ and $\ro-R={\cal O}(1/\xi^{3/2})$. 
Specifically, solving (\ref{r3ssss1})--(\ref{r3ssss2}) we find 
\beq
N \sim \sqrt{\frac{\beta\ro}{\xi}} , \qquad 
R \sim \ro - \frac{4\mu\nu}{\alpha\ro} \sqrt{\frac{\beta\ro}{\xi}} , 
\lbl{r3a2-sss} \eeq
hence the dimer concentrations $c = \half (\ro-R) \sim N^3 = 
{\cal O}(1/\xi^{3/2})$ and $z = 2 N^2/\ro \sim N^2 = {\cal O}(1/\xi)$. 
More precisely, $c\sim (2\mu\nu/\alpha)\sqrt{\beta/\ro\xi}$ and 
$z\sim 2\beta/\xi$, in contrast with the previous asymptotic scaling 
which gave $z\sim N^2$). 

To determine the timescales for crystal growth and dissolution, 
we use (\ref{r3a2-sss}) to define %% new stuff
\beq 
N \sim n(t) \sqrt{\beta \ro/\xi} , \quad 
R \sim \ro - \frac{4\mu\nu r(t)}{\alpha \ro} 
\sqrt{\frac{\beta\ro}{\xi}} , 
\eeq 
and so rewrite the governing equations (\ref{r3Rd})--(\ref{r3Nd}) as 
\beqa
\frac{\dd n}{\dd t} & = & \beta n \left( 1 - n^2 - 
\frac{2 n (\beta+\mu\nu)}{\sqrt{\ro\xi\beta}} \right) , \\ 
\frac{\dd r}{\dd t} & = & \alpha \sqrt{\frac{\beta\ro}{\xi}} 
\left( n^2 -r - \frac{2\mu r}{\alpha} 
\sqrt{\frac{\xi}{\beta\ro}} \right) . 
\eeqa
Here, the former equation for $n(t)$ corresponds to the 
slower timescale, with a rate $\beta$, the rate of 
equilibration of $r(t)$ being $\alpha \sqrt{\beta\ro/\xi}$. 

The stability of the symmetric state is determined by 
\beq
\frac{R}{N} \frac{\dd }{\dd t} 
\left( \begin{array}{c} \phi(t) \\ \zeta(t) \end{array} \right) = 
\left( \begin{array}{cc} -2 \sqrt{\beta\ro\xi} & \sqrt{\beta\ro\xi} \\ 
-4\mu\nu \sqrt{\beta / \xi \ro} & 4\mu\nu \end{array} \right) 
\left( \begin{array}{c} \phi \\ \zeta \end{array} \right) . 
\lbl{r3a2-phi-zeta-sys} \eeq 
This matrix has one large negative eigenvalue ($\sim -2\sqrt{\beta\ro\xi}$) 
and one (smaller) positive eigenvalue ($\sim 4\mu\nu$); the former 
corresponds to $(1,0)^T$ hence the decay of $\phi$, whilst the latter 
corresponds to the eigenvector $(1,2)^T$.  Hence the system 
(\ref{r3a2-phi-zeta-sys}) has the solution 
\beq 
\left( \begin{array}{c} \phi \\ \zeta \end{array} \right) \sim 
C \left( \begin{array}{c} 1 \\ 2 \end{array} \right) 
\exp \left( 4 \mu \nu t \sqrt{ \frac{\beta}{\ro\xi}} \right) . 
\lbl{r3a2urate} \eeq 
The chiralities evolve on two timescales, the faster being 
$2\beta$ corresponding to the stable eigenvalue of 
(\ref{r3a2-phi-zeta-sys}) and the slower unstable rate 
being $4\mu\nu\sqrt{\beta/\xi\ro}$.  This timescale is 
similar to (\ref{r3-chir-rate}), being dependent on mass and 
the ratio of aggregation to fragmentation, and inversely 
proportional to the chiral switching rate of dimers ($\mu\nu$).  

\begin{figure}[!ht]
\vspace*{68mm}
\includegraphics{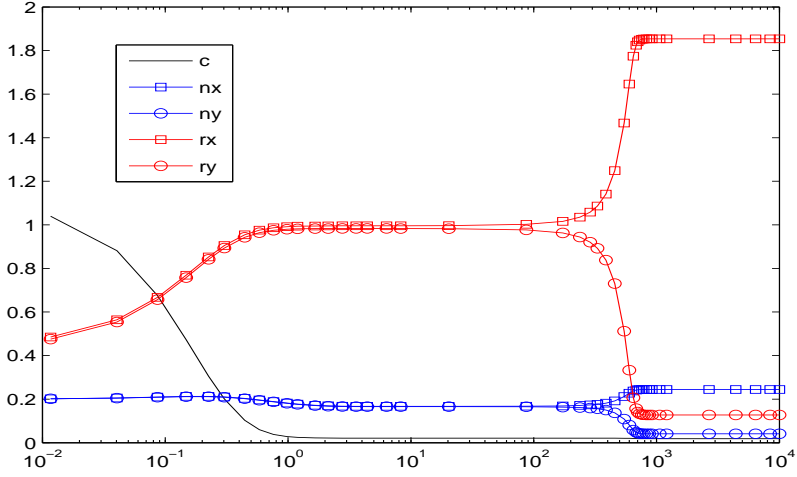}
\caption{Graph of the concentrations $N_x,N_y,\ro_x,\ro_y,c$ 
against time on a logarithmic time for the asymptotic limit 2, 
with initial conditions $N_x=0.2=N_y$, $\ro_x=0.45$, $\ro_y=0.44$, 
other parameters given by $\alpha=10=\xi$, $\beta=1=\mu$, 
$\nu=0.5$, $\ro=2$. Since model equations 
are in nondimensional form, the time units are arbitrary. }
\label{fig-r3beta}
\end{figure}

%----------------------
\subsection{The asymmetric steady-state}

Since the symmetric state can be unstable, there must be some 
other large-time asymmetric attractor(s) for the system, 
which we now aim to find.  From (\ref{r3-nxdot}) and 
(\ref{r3-roxdot}), at steady-state, we have 
\beq
2c_2 (2\mu+\alpha N_x) = \frac{4\mu\nu N_x^2}{\ro_x} , 
\qquad \mu c_2 + \beta N_x = 
2 (\mu\nu+\beta+\xi N_x) \frac{N_x^2}{\ro_x} . 
\lbl{r3a-eqs} \eeq 
Taking the ratio of these we find a single quadratic 
equation for $N_x$
\beq
0 = \alpha \xi N_x^2 - \left( \frac{\beta\mu\nu}{c_2} 
- \alpha\beta - \alpha\mu\nu - \xi\mu \right) N_x + \beta\mu , 
\lbl{r3a-Neq}
\eeq
with an identical one for $N_y$.   Hence there is the possibility 
of distinct solutions for $N_x$ and $N_y$ if both roots of 
(\ref{r3a-Neq}) are positive; this occurs if 
\beq
c_2 < \frac{\beta\mu\nu}{\alpha\beta + \xi\mu 
+ \alpha\mu\nu + 2\sqrt{\alpha\beta\xi\mu} } . 
\lbl{r3a-ineq} \eeq
Given $N_x$ ($N_y$), we then have to solve one of 
(\ref{r3a-eqs}) to find $\ro_x$ ($\ro_y$), {\em via} 
\beq
\ro_x = \frac{2 \mu \nu N_x^2}{c_2 (\mu+\alpha N_x)} ,  
\lbl{r3a-a1-rox} \eeq 
and then satisfy the consistency condition that 
$\ro_x + \ro_y + 2 c_2 = \ro$.  After some algebra, 
this condition reduces to 
\beqa
\half \alpha^2 \xi c_2^2 (\beta \!-\! \alpha c_2 ) (\ro\!-\!2c_2) 
&\!=\!& \beta^2\mu^2\nu^2 - \beta\mu\nu c_2 [ \alpha\beta 
+ 2\alpha\mu\nu + 2\xi\mu ] \nn \\ && 
+ \mu c_2^2 [ \mu (\alpha\nu\!+\!\xi)^2 + \alpha\beta (\alpha\nu\!-\!\xi) ] . 
\lbl{r3a-consistency} \eeqa
Being a cubic, it is not straightforward to write down explicit 
solutions of this equation, hence we once again consider the two 
asymptotic limits ($\beta\ll1$ and $\alpha\sim\xi\gg1$).  

%------------------
\subsubsection{Asymptotic limit 1: $\beta \ll 1$}

In this case, $c_2 = {\cal O}(\beta)$ hence we put $c_2=\beta C$ 
and the consistency condition (\ref{r3a-consistency}) yields  
\beq 
{\cal O}(\beta^3) = \beta^2 \left[ \nu - (\alpha\nu+\xi) C \right]^2 , 
\lbl{r3a-a1-C} \eeq 
hence, to leading order, $C=\nu/(\alpha\nu+\xi)$ . Unfortunately, the 
resulting value for $c_2$ leads to all the leading order terms in the 
linear equation (\ref{r3a-Neq}) for $N_x$ to cancel. We thus have to 
find higher order terms in the expansion for $c_2$; due to the form of 
(\ref{r3a-a1-C}), the next correction term is ${\cal O}(\beta^{3/2})$. 
Putting $c_2=\beta C(1+\tilde C \sqrt{\beta})$, we find 
\beq 
\tilde C^2 = \frac{\alpha\xi \,\left[ \, \alpha\xi\ro + 4 \mu (\alpha\nu+\xi) 
\, \right] }{2\mu^2 (\alpha\nu+\xi)^3} . 
\eeq 
In order to satisfy the inequality (\ref{r3a-ineq}), we require the 
negative root, that is, $\tilde C<0$. 

Although the formulae for $N_x,N_y$ are lengthy, 
their sum and products simplify to 
\beq
\Sigma = N_x + N_y = 
\frac{\mu \tilde C \sqrt{\beta} (\alpha\nu+\xi)}{\alpha\xi} , \qquad 
\Pi = N_x N_y = \frac{\beta\mu}{\alpha\xi} . 
\eeq
The chirality $\phi$ can be simplified using $\phi^2=1-4\Pi/\Sigma^2$ 
which implies  
\beq
\phi^2 = \frac{\alpha\ro \xi - 4\mu(\alpha\nu+\xi)}
{\alpha\ro\xi+4\mu (\alpha\nu+\xi)}  . 
\eeq
Hence we require $\ro > \ro_c := 4\mu(\alpha\nu+\xi)/\alpha\xi$ in 
order for the system to have nonsymmetric steady-states, that is, the 
system undergoes a symmetry-breaking bifurcation as $\ro$ increases 
through $\ro=\ro_c$.  As the mass in the system increases further, 
the chirality $\phi$ approaches ($\pm$) unity, indicating a state in 
which one handedness of crystal completely dominates the other. 

%------------------
\subsubsection{Asymptotic limit 2: $\alpha \sim \xi \gg 1$}

In this case,  the left-hand side of the consistency condition 
(\ref{r3a-consistency}) is ${\cal O}(\alpha^2\xi c_2^2)$ whilst the 
right-hand side is ${\cal O}(1)+{\cal O}(\alpha c_2^2)$, which implies 
the balance $c_2={\cal O}(\xi^{-3/2})$.  Solving for $c_2$ leads to 
\beq 
c_2 \sim \frac{\mu\nu}{\alpha} \sqrt{ \frac{2\beta}{\ro\xi} } . 
\eeq 
The leading order equation for $N_x,N_y$ is then 
\beq 
0 = \alpha\xi N^2 - \alpha N \sqrt{\half\beta\ro\xi} + \beta\mu , 
\eeq 
hence we find the roots 
\beq 
N_x,N_y \sim \sqrt{\frac{\beta\ro}{2\xi}} , 
\frac{2\mu}{\alpha} \sqrt{\frac{\beta}{2\xi\ro}} , \qquad 
\ro_x , \ro_y \sim \ro , \frac{2\mu}{\alpha} . 
\eeq 
Since we have either $\ro_x \gg N_x \gg \ro_y \gg N_y$ 
or $\ro_y \gg N_y \gg \ro_x \gg N_x$, in this asymptotic limit, 
the system is completely dominated by one species or the other. 
Putting $\Sigma=N_x+N_y$ and $\Pi=N_xN_y$ we have 
$\phi^2=1-4\Pi/\Sigma^2 \sim 1 - 8 \mu/\alpha\ro$.  

%--------------------------------------------------------
% 5 ESTIMATION OF RATE OF CHIRAL AMPLIFICATION %%
\section{Discussion}
\label{disc-sec}
\setcounter{equation}{0}

We now try to use the above theory and experimental 
results of Viedma \cite{viedma} to estimate the relevant 
timescales for symmetry-breaking in a prebiotic world. 
Extrapolating the data of time against grinding rate in rpm 
from Figure 2 of Viedma \cite{viedma} 
suggests times of $2\times10^5$ hours using a straight line 
fit to log(time) against log(rpm) or 1000--3000 hours if 
log(time) against rpm or time against log(rpm) is fitted. 
A reduction in the speed of grinding in prebiotic circumstances 
is expected since natural processes such as water waves 
are much more likely to operate at the order of a few seconds$^{-1}$ 
or minutes$^{-1}$ rather than 600 rpm.   

Similar extrapolations on the number and mass of balls 
used to much lower amounts gives a further reduction 
of about 3, using a linear fit to log(time) against mass of balls 
from Figure 1 of Viedma \cite{viedma}.   There is an 
equally good straight line fit to time against log(ball-mass) 
but it is then difficult to know how small a mass of balls 
would be appropriate in the prebiotic scenario. 
There is an additional factor due to the experiments 
of Viedma being on a small volume of 10 ml, whereas a 
sensible volume for prebiotic chemistry is 1000 l, 
giving an additional factor of $10^5$. 
Combining these three factors ($10^3$, 3, and $10^5$) with 
the 10 days of the original experiment, we estimate that the 
timescale for prebiotic symmetry breaking is ${\cal O}(3\times10^9)$ 
days, which is equivalent to the order of about ten million years.  

This extrapolation ignores the time required to arrive 
at the initial enantiomeric excesses of 5\% used by Viedma 
\cite{viedma} from a small asymmetry caused by 
either a random fluctuation or by the parity-violation.   
Although the observed chiral structures are the minimum energy 
configurations as predicted by parity violation, there is an evens 
probability that the observed handedness could simply be the result 
of a random fluctuation which was amplified by the same mechanisms. 
In order to perform an example calculation, we take a random 
fluctuation of the size predicted by parity violation, which is 
of the order of $10^{-17}$, as suggested by Kondepudi \& Nelson 
\cite{kon-pla}.  Our goal is now to find the time taken to 
amplify this to an ${\cal O}(1)$ (5\%) enantiomeric excess. 

The models derived in this paper, for example in Section 
\ref{r3-a2-sec}, predict that the chiral excess grows 
exponentially in time.  Assuming, from (\ref{r3a2urate}), 
that $\phi(t_0)=10^{-17}$ and $\phi(t_1)= 0.1$, then 
the timescale for the growth of this small perturbation is 
\[ t_1 - t_0 = \frac{1}{4\mu\nu} \sqrt{\frac{\xi\ro}{\beta}} 
\log \frac{10^{-1}}{10^{-17}} . \] 
Since the growth of enantiomeric excess is exponential, 
it only takes 16 times as long for the perturbation to grow 
from $10^{-17}$ to $10^{-1}$ as from $10^{-1}$ to 1. 
Hence we only need to increase our estimate of the timescale 
by one power of ten, to 100 million years. 

This estimate should be taken as a very rough estimate, since it 
relies on extrapolating results by many orders of magnitude. 
Also, given the vast differences in temperature from the 
putative subzero prebiotic world to a tentative hot hydrothermal 
vent, there could easily be changes in timescale by a factor of 
several orders of magnitude. 

%--------------------------------------------------------
\section{Conclusions}
\label{conc-sec}
\setcounter{equation}{0}

After summarising the existing models of chiral symmetry-breaking 
processes we have systematically derived a model in which through 
aggregation and fragmentation chiral clusters compete for achiral 
material.  The model is closed, in that there is no input of mass 
into the system, although the form of the aggregation and 
fragmentation rate coefficients mean that there is an input of energy, 
keeping the system away from equilibrium. Furthermore, there is no 
direct interaction of clusters of opposite handedness; rather 
just through a simple competition for achiral substrate, the system 
can spontaneously undergo chiral symmetry-breaking.  This model 
helps explain the experimental results of Viedma \cite{viedma} 
and Noorduin \etal\ \cite{wim}. 

The microscopic model originally derived has been simplified 
successively to a minimalistic model, which, numerical results show, 
exhibits symmetry-breaking. Even after this reduction, the model is 
extremely complex to analyse due to the large number of cluster 
sizes retained in the model. Hence we construct two truncated 
models, one truncated at tetramers, which shows no 
symmetry-breaking and one at hexamers which shows 
symmetry-breaking under certain conditions on the parameter values.  
Alternative reductions are proposed: instead of retaining the 
concentrations of just a few cluster sizes, we retain information 
about the shape of the distribution, such as the number of clusters 
and the total mass of material in clusters of each handedness.  
These reduced models are as simple to analyse as truncated models 
yet, since they more accurately account for the shape of the 
size-distribution than a truncated model, are expected to give models 
which more easily fit to experimental data.  Of course, other 
ansatzes for the shape of the size distributions could be made, 
and will lead to modified conditions for symmetry-breaking; 
however, we believe that the qualitative results outlined here will 
not be contradicted by analyses of other macroscopic reductions. 

One noteworthy feature of the results shown herein is that the 
symmetry-breaking is inherently a product of the two handednesses 
competing for achiral material.  The symmetry-breaking does not rely 
on critical cluster sizes, which are a common feature of theories of 
crystallisation, or on complicated arguments about surface area to 
volume ratios to make the symmetric state unstable.  We do not 
deny that these aspects of crystallisation are genuine, these features 
are present in the phenomena of crystal growth, but they are not 
the fundamental cause of chiral symmetry-breaking. 

More accurate fitting of the models to experimental data could be 
acheived if one were to fit the generalised Becker-D\"{o}ring model 
(\ref{gbd1})--(\ref{gbd3}) with realistic rate coefficients.  Questions 
to address include elucidating how the number and size distribution 
at the start of the grinding influences the end state.  For example, if 
one were to start with a few large right-handed crystals and many 
small left-handed crystals, would the system convert to entirely 
left- or entirely right-handed crystals ?  Answers to these more 
complex questions may rely on higher moments of the size distributions, 
surface area to volume ratios and critical cluster nuclei sizes. 

%----------------------------------
\subsection*{Acknowledgments}

I would particularly like to thank Professors Axel Brandenburg and 
Raphael Plasson for inviting me to an extended programme of 
study on homochirality at Nordita (Stockholm, Sweden) in February 2008.  
There I met and benefited greatly from discussions with Professors 
Meir Lahav, Mike McBride, Wim Noorduin, as well as many others.  
The models described here are a product of the stimulating 
discussions held there.  I am also grateful for funding under 
EPSRC springboard fellowship EP/E032362/1. 

%----------------------------------
\appendix
\renewcommand{\theequation}{\Alph{section}\arabic{equation}}
\section{General theory for crystallisation and grinding 
with competition between polymorphs} 
\label{app}
\setcounter{equation}{0}

This model can be generalised so as to be applicable to the case 
of grinding a system undergoing crystallisation in which several 
polymorphs of crystal nucleate simultaneously.  It may then be 
possible to use grinding to suppress the growth of one polymorph 
and allow a less stable form to be expressed.  In this case, the 
growth and fragmentation rates of the two polymorphs will differ, 
we denote the two polymorphs by $x$ and $y$.  In place of $a$, 
$b$, $\alpha$, $\xi$, $\beta$ we have $a_{x,r}$, $a_{y,r}$, $b_{x,r}$, 
$\alpha_{x,r}$, etc.   Hence in place of (\ref{gbd-c1})--(\ref{gbd-y2}) 
we have 
\beqa
\frac{\dd x_r}{\dd t} \!&\!=\!& \!a_{x,r-1}c_1x_{r-1} \!-\! b_{x,r} x_r 
	\!-\! a_{x,r} c_1 x_r \!+\! b_{x,r+1} x_{r+1} 	
	\!-\! \beta_{x,r} x_r \!+\! \beta_{x,r+2} x_{r+2} \nn \\ && 
	\!+ (\alpha_{x,r-2} c_2 \!+\! \xi_{x,r-2} x_2 ) x_{r-2} 
	\!-\! (\alpha_{x,r} c_2 \!+\! \xi_{x,r} x_2) x_{r}, \quad (r\geq4) , \\ 
\frac{\dd y_r}{\dd t} \!&\!=\!&\! a_{y,r-1} c_1 y_{r-1} \!-\! b_{y,r} y_r 
	\!-\! a_{y,r} c_1 y_r \!+\! b_{y,r+1} y_{r+1} 
	\!-\! \beta_{y,r} y_r \!+\! \beta_{y,r+2} y_{r+2} \nn \\ && 
	\!+ (\alpha_{y,r-2} c_2 \!+\! \xi_{y,r-2} y_2) y_{r-2} 
	\!-\! (\alpha_{y,r} c_2 \!+\! \xi_{y,r} y_2) y_{r} , \quad (r\geq4) , \\ 
\frac{\dd x_2}{\dd t} \!&\!=\!& \mu_x c_2 - \mu_x \nu_x x_2 
	- a_{x,2} c_1 x_2 + b_{x,3} x_3 
	- (\alpha_{x,r} c_2 + \xi_{x,r} x_2) x_r \nn \\ && 
	+ \beta_{x,4} x_4 + \sum_{k=4}^\infty \beta_{x,r} x_r   
	- \sum_{k=2}^\infty \xi_{x,k} x_2 x_k , \\ 
\frac{\dd y_2}{\dd t} \!&\!=\!& \mu_y c_2 - \mu_y \nu_y y_2 
	- a_{y,2} c_1 y_2 + b_{y,3} y_3 
	- (\alpha_{y,r} c_2 + \xi_{y,r} y_2) y_{r} \nn \\ && 
	+ \beta_{y,4} y_4 + \sum_{k=4}^\infty \beta_{y,r} y_r   
	- \sum_{k=2}^\infty \xi_{y,k} y_2 y_k , \\ 
\frac{\dd x_3}{\dd t} \!&\!=\!&\! a_{x,2} x_2 c_1 \!-\! b_{x,3} x_3 
	\!-\! a_{x,3 }c_1 x_3 \!+\! b_{x,4} x_4 
	\!-\! (\alpha_{x,3} c_2 \!+\! \xi_{x,3} x_2) x_3 
	\!+\! \beta_{x,5} x_5 , \nn \\ && \\ 
\frac{\dd y_3}{\dd t} \!&\!=\!&\! a_{y,2} y_2 c_1 \!-\! b_{y,3} y_3 
	\!-\! a_{y,3} c_1 y_3 \!+\! b_{y,4} y_4 
	\!-\! (\alpha_{y,3} c_2 \!+\! \xi_{y,3} y_2) y_3 
	\!+\! \beta_{y,5} y_5 , \nn \\ && \\
\frac{\dd c_2}{\dd t} \!&\!=\!&\! \mu_x \nu_x x_2 \!+\! \mu_y \nu_y y_2 
	\!-\! (\mu_x\!\!+\!\!\mu_y) c_2 \!\!+\!\! \delta c_1^2 \!\!-\!\! \epsilon c_2 
	\!\!-\!\! \sum_{k=2}^\infty 
	\! c_2 ( \alpha_{x,r} x_r \!\!+\!\! \alpha_{y,r} y_r ) , \\   
\frac{\dd c_1}{\dd t} \!&\!=\!&\! 2 \epsilon c_2 \!-\! 2\delta c_1^2 
	\!-\!\sum_{k=2}^\infty ( a_{x,k} c_1 x_k \!-\! b_{x,k+1} x_{k+1} 
	\!+\! a_{y,k} c_1 y_k \!-\! b_{y,k+1} y_{k+1} ) . \nn \\ && 
\eeqa 

For simplicity let us consider an example in which all the growth 
and fragmentation rate parameters are independent of cluster size, 
($a_{x,r}=a_x$, $\xi_{y,r}=\xi_y$, etc.~for all $r$). The thermodynamic 
stability of the two types of crystal depends on their relative 
interactions with monomers from solution, that is, if $a_x/b_x > 
a_y/b_y$ then $X$ is the more stable form. This is because, in the 
absence of $c_2$, we can define free energy functions 
\beq 
Q^{x}_r = \left( \frac{a_x}{b_x} \right)^{r-1} , \qquad 
Q^{y}_r = \left( \frac{a_y}{b_y} \right)^{r-1} , 
\eeq 
which generate the equilibrium distributions 
\beq  
c_r^{eqx} = Q_r^{x} c_1^r = \frac{b_x}{a_x} 
\left( \frac{a_x c_1}{b_x} \right)^r   \;\; > \;\;  
c_r^{eqy} = Q_r^{y} c_1^r = \frac{b_y}{a_y} 
\left( \frac{a_y c_1}{b_y} \right)^r . 
\eeq 
If $a_x/b_x<a_y/b_y$ then the latter ($Y$) will be the dominant 
crystal type {\sl at equilibrium}, whilst $X$ is the less stable 
morphology at equilibrium. These last two words are vital, since, 
at early times, the growth rates depend on the relative sizes of the 
growth rates $a_x$ and $a_y$.   It is possible for the less stable form 
to grow first and more quickly from solution, and be observed for a 
significant period of time, since the rate of convergence to equilibrium 
also depends on the fragmentation rates and so can be extremely 
slow (see Wattis \cite{jw-comp} for details). 

In the presence of grinding, the crystal size distributions also 
depend upon the strength of dimer interactions, that is, the growth 
rates $\alpha_x c_2\!+\!\xi_x x_2$, $\alpha_y c_2 \!+\! \xi_y y_2$ 
and the grinding rates $\beta_x$, $\beta_y$.  The steady-state size 
distributions will depend on the relative growth ratios due to 
grinding $(\alpha_x c_2 \!+\! \xi_x x_2)/\beta_x$ and 
$(\alpha_y c_2 \!+\! \xi_y y_2)/\beta_y$ as well as the more traditional 
terms due to growth from solution, namely $a_x c_1/b_x$ and 
$a_y c_1/b_y$.  Such systems with dimer interactions 
have been analysed previously by Bolton \& Wattis 
\cite{bw-dimers}. The presence of dimer interactions can 
alter the size distribution, and in non-symmetric systems 
such as those analysed here, dimer interactions can alter the 
two distributions differently. Two points are worth noting here: 
\\ \phantom{.}\quad 
(i) for certain parameter values, the less stable stable form ($Y$, say, 
with $a_y/b_y < a_x/b_x$) may be promoted to the more stable 
morphology by grinding (if $(\alpha_y c_2 \!+\! \xi_y y_2) / \beta_y$ 
is sufficiently greater than $(\alpha_x c_2 \!+\! \xi_x x_2) / \beta_x$); 
\\ \phantom{.}\quad 
(ii) grinding may make a less rapidly nucleating and growing form 
($Y$, say, with $a_y < a_x$) into a more rapidly growing form if 
$\alpha_y c_2 \!+\! \xi_y y_2$ is sufficiently greater than 
$\alpha_x c_2 \!+\! \xi _2 x_2$. 

In systems which can crystallise into three or more forms, we may 
have the case where $x$ is more stable than $y$ and $y$ is more 
stable than $z$; thus, at equilibrium $x$ will be observed.  
Furthermore, if $a_x <  a_y > a_z$ we may observe type $y$ at 
early times due to it having faster nucleation and growth rates than 
$x$ and $z$.   
However, it is possible that the presence of grinding could 
suppress both $x$ and $y$ and allow $z$ to be expressed, 
if some combination of the inequalities 
\beq 
\frac{\alpha_z c_2+\xi_z z_2}{\beta_z} \; > \; 
\frac{\alpha_y c_2+\xi_y y_2}{\beta_y} \; , \;\;
\frac{\alpha_x c_2+\xi_x x_2}{\beta_x} , 
\eeq  
$\alpha_z>\alpha_y,\alpha_x$, $\xi_z>\xi_x,\xi_y$ hold. 

%-------------------------------------------------------------


\footnotesize

\begin{thebibliography}{99}
\parsep=0pt
\itemsep=0pt

\bibitem{bd} R Becker, W D\"{o}ring. Kinetische behandlung der 
keimbildung in \"{u}bers\"{a}ttigten d\"{a}mpfen. {\em Ann Phys}, 
{\bf 24}, 719--752, (1935)

\bibitem{jw-polymorph} CD Bolton \& JAD Wattis. The Becker-D\"{o}ring 
equations with input, competition and inhibition. {\em J Phys A; Math 
Gen}, {\bf 37}, 1971--1986, (2004). 

\bibitem{bw-dimers} CD Bolton \& JAD Wattis. Generalised 
Becker-D\"{o}ring equations:  effect of dimer interactions. 
{\em J Phys A; Math Gen}, {\bf 35}, 3183--3202, (2002). 

\bibitem{bw} CD Bolton \& JAD Wattis. Generalised coarse-grained 
Becker-D\"{o}ring equations. {\em J Phys A; Math Gen}, {\bf 36}, 7859--7888, (2003). 

\bibitem{axel} A Brandenburg, AC Andersen, S H\"{o}fner, M Nilsson.  
Homochiral growth through enantiomeric cross-inhibition. {\em Origins of Life 
and Evolution of Biospheres}, {\bf 35}, 225--241, (2005). {\tt arXiv:q-bio/0401036}. 

\bibitem{axel3} A Brandenburg, AC Andersen, M Nilsson.   Dissociation in a 
polymerization model of homochirality.  {\em Origins of Life and Evolution 
of  Biospheres}, {\bf 35}, 507--521, (2005). {\tt arXiv:q-bio/0502008}

\bibitem{fern}  FP da Costa. Asymptotic behaviour of low density 
solutions to the generalized Becker-D\"{o}ring equations. 
{\em Nonlinear Diff Eq Appl}, {\bf 5}, 23--37, (1998). 

\bibitem{childs} PV Coveney, JAD Wattis. Coarse-graining and renormalisation 
group methods for the elucidation of the kinetics of complex nucleation and 
growth processes. {\em Mol Phys}, {\bf 104}, 177--185, (2006). 

\bibitem{darwin} C Darwin. Private letter to Joseph Hooker, (1871).  
Published in (pp.168--169 of) F Darwin (ed), The life and letters of 
Charles Darwin, including an autobiographical chapter, 3 vols. 
John Murray, London, (1887). 

\bibitem{frank} FC Frank.  On spontaneous asymmetric synthesis. 
{\em Biochim Biophys Acta}, {\bf 11}, 459--463, (1953).  

\bibitem{sara-bup}  M Gleiser \& SI Walker.  An extended model for 
the evolution of prebiotic homochirality: a bottom-up approach to 
the origin of life. {\tt arXiv.org/0802.2884} [q-bio.BM], (2008). 

\bibitem{sara-punc}  M Gleiser, J Thorarinson \& SI Walker.  Punctuated 
Chirality.  {\tt arXiv.org/0802.1446} [astro-ph], (2008). 

\bibitem{kon+a}  DK Kondepudi \& K Asakura. Chiral autocatalysis, 
spontaneous symmetry breaking and stochastic behaviour. 
{\em Acc Chem Res}, {\bf 34}, 946--954, (2001). 

\bibitem{kon-jacs}  DK Kondepudi, KL Bullock, JA Digits PD Yarborough. 
Stirring rate as a critical parameter in chiral symmetry breaking 
crystallization. {\em J Am Chem Soc}, {\bf 117}, 401--404, (1995). 

\bibitem{kon-sci} DK Kondepudi, RJ Kaufman \& N Singh. 
Chiral symmetry-breaking in sodium chlorate crystallization. 
{\em Science}, {\bf 250}, 975--976, (16 Nov, 1990). 

\bibitem{kon-pla} DK Kondepudi \& GW Nelson.  Chiral symmetry breaking 
in nonequilibrium chemical systems: time scales for chiral selection.  
{\em Phys Lett A}, {\bf 106}, 203--206, (1984). 

\bibitem{kon-nat} DK Kondepudi \& GW Nelson.   Weak neutral currents 
and the origin of biomolecular chirality.  {\em Nature}, {\bf 314}, 438--441, (1985).

\bibitem{mcbride-nature} JM McBride \& JC Tully.  Did life grind to a start? 
{\em Nature (News and views)}, {\bf 452}, 161--162, (13 March, 2008).   

\bibitem{axel2} T Multamaki, A Brandenburg. Spatial dynamics of homochiralization.  
{\em Int J Astrobiol}, {\bf4}, 73--78, (2005). {\tt arXiv:q-bio/0505040}. 

\bibitem{Murray} JD Murray. Mathematical Biology  (Biomathematics, 
vol 19). Springer-Verlag, Berlin, (1989). [Particularly App A2.1, pp.702--703] 

\bibitem{wim} WL Noorduin, T Izumi, A Millemaggi, M Leeman, H Meekes, 
WJP van Enckevort, RM Kellogg, B Kaptein, E Vlieg \& DG Blackmond.   
Emergence of a single solid chiral state from a nearly racemic amino 
acid derivative.  {\em J Am Chem Soc}, {\bf 130}, 1158--1159, (2008). 

\bibitem{plasson} R Plasson, H Bersini \& A Commeyras.  Recycling 
Frank: spontaneous emergence of homochirality in noncatalytic systems.  
{\em Proc Natl Acad Sci}, {\bf 101}, 16733--16738, (2004).  

\bibitem{saito} Y Saito \& H Hyuga.   Complete homochirality induced by 
the nonlinear autocatalysis and recycling.   {\em J Phys Soc Jap}, {\bf 
73}, 33--35, (2004). Also available at {\tt arXiv:physics/0310142} 

\bibitem{saito2} Y Saito \& H Hyuga. Chirality selection in crystallization. 
{\em J Phys Soc Jpn}, {\bf  74},  535--537, (2005). 

\bibitem{sandars} PGH Sandars. A toy model for the generation of 
homochirality during polymerisation.  {\em Origins of Life and Evolution 
of Biospheres}, {\bf 33}, 575--583, (2003).

\bibitem{smol}  M von Smoluchowski. Drei vortr\"{a}ge \"{u}ber 
diffusion Brownsche molekular bewegung und koagulation von 
kolloidteichen. {\em Physik Z}, {\bf 17}, 557--571, (1916). 

\bibitem{soai} K Soai, T Shibata, H Morioka \& K Choji. Asymmetric 
autocatalysis and amplification of enantiomeric excess of a 
chiral molecule. {\em Nature}, {\bf 378}, 767--768, (1995).  

\bibitem{uwaha} M Uwaha. A model for complete chiral crystallization. 
{\em J Phys Soc Jap}, {\bf 73}, 2601--2603, (2004). 

\bibitem{viedma} C Viedma.   Chiral symmetry breaking during 
crystallization: complete chiral purity induced by nonlinear autocatalysis 
and recycling. {\em Phys Rev Lett}, {\bf 94}, 065504, (2005). 

\bibitem{jw-comp} JAD Wattis. A Becker-D\"{o}ring model of competitive 
nucleation.  {\em J Phys A: Math Gen}, {\bf 32}, 8755--8784, (1999). 

\bibitem{jw-cement} JAD Wattis \& PV Coveney. Generalised nucleation 
theory with inhibition for chemically reacting systems. 
{\em J Chem Phys}, {\bf 106}, 9122--9140, (1997). 

\bibitem{jw-sandars} JAD Wattis \& PV Coveney. Symmetry-breaking 
in chiral polymerisation,  {\em Origins of Life and the evolution of 
Biospheres}, {\bf 35}, 243--273, (2005). 

\bibitem{jw-ch-rna-rev} JAD Wattis \& PV Coveney. Chiral polymerisation 
and the RNA world, {\em Int J Astronomy}, {\bf 4}, 63--73, (2005). 

\end{thebibliography}
\end{document}